\documentclass[12pt]{article}

\usepackage{times}
\usepackage{citesort}
\usepackage{epsfig}
 \psfull

\newcommand{\comment}[1]{}
\usepackage{url}
\usepackage{amssymb}
\usepackage{amsbsy}
\usepackage{amsmath}
\usepackage{graphicx}
\usepackage{color}

\newcommand{\mathReal}{\mathbb{R}}
\newcommand{\mathComplex}{\mathbb{C}}

\newcommand{\Sphere}[1]{\mathbf{S}^{#1}}

\newcommand{\OO}[1]{\mathbf{O}({#1})}

\newcommand{\PGL}[1]{\mathbf{PGL}({#1})}
\newcommand{\GL}[1]{\mathbf{GL}({#1})}
\newcommand{\RP}[1]{\mathbb{R}\mathbf{P}^{#1}}
\newcommand{\CP}[1]{\mathbb{C}\mathbf{P}^{#1}}

\begin{document}


\title{A Contour Integral Representation for the
Dual Five-Point Function and a Symmetry of the Genus Four Surface in $\mathReal^{6}$}

\author{Andrew J.~Hanson  and Ji-Ping Sha
  \thanks{Email: \{hansona, jsha\}@indiana.edu} \\
  Computer Science Dept.  and Mathematics Dept. \\[.0in]
  Indiana University  \\
  Bloomington, IN \ 47405 \ USA   }

\maketitle

\begin{abstract}
  The invention of the ``dual resonance model'' $N$-point functions
  $B_{N}$ motivated the development of current string theory.  The
  simplest of these models, the four-point function $B_{4}$, is the
  classical Euler Beta function.  Many standard methods of complex
  analysis in a single variable have been applied to elucidate the
  properties of the Euler Beta function, leading, for example, to
  analytic continuation formulas such as the contour-integral
  representation obtained by Pochhammer in 1890.  However, the precise
  features of the expected multiple-complex-variable generalizations to
  $B_{N}$ have not been systematically studied.  Here we
  explore the geometry underlying the dual five-point function
  $B_{5}$, the simplest generalization of the Euler Beta function.
  The original integrand defining $B_{5}$ leads to a polyhedral
  structure for the five-crosscap surface, embedded in $\RP{5}$, that
  has 12 pentagonal faces and a symmetry group of order 120 in
  $\PGL{6}$.  We find a Pochhammer-like representation for $B_{5}$
  that is a contour integral along a surface of genus five in $\CP{2}
  \# 4\overline{\CP{2}}$.  The symmetric embedding of the
  five-crosscap surface in $\RP{5}$ is doubly covered by a
  corresponding symmetric embedding of the surface of genus four in
  $\Sphere{5} \subset \mathReal^{6}$ that has a polyhedral structure with 24
  pentagonal faces and a symmetry group of order 240  in
  $\OO{6}$.  These symmetries enable the construction of elegant
  visualizations of these surfaces.  The key idea of the paper is to
  realize that the compactification of the set of five-point
  cross-ratios forms a smooth real algebraic subvariety that is the
  five-crosscap surface in $\RP{5}$.  It is in the complexification of
  this surface that we construct the contour integral representation
  for $B_{5}$.  Our methods are generalizable in principle to higher
  dimensions, and therefore should be of interest for further study.
\end{abstract}

\section{Introduction}
\label{introduction.sec}

\paragraph{Historical Background.}
In 1968, Gabriele Veneziano \cite{veneziano1968} noticed that an
amazing number of abstract properties required by the relativistic
scattering amplitude for four colliding spinless particles were
embodied in the classical Euler Beta function,
$B(\alpha_{1},\alpha_{2})$, which can be 
defined by the integral representation
\begin{equation}
B(\alpha_{1},\alpha_{2}) =
\int_{0}^{1}\,x^{\alpha_{1}-1}(1-x)^{\alpha_{2}-1}
\, dx  \ , \ \ {\rm Re}\,\alpha_{1} >0, \, {\rm Re}\,\alpha_{2} >0   \ .
\label{B4.eq}
\end{equation}
This observation served as the implausible origin of modern string
theory (see, e.g., \cite{polchinksyI,polchinksyII}
for more details), which grew
from the discovery that the Beta function could be
related to the vibration modes of a relativistic string sweeping
out a surface in spacetime \cite{nambu1970,goto1971}.

Almost immediately following Veneziano's discovery, a
function  with a two-dimensional integral representation was found that
could be related to the relativistic scattering amplitude of {\it
five\/} spinless particles \cite{BardakciRuegg1968,Virasoro1969}.
This function, the dual five-point  function $B_{5}$, can be written in various
representations such as the following integral over a
triangular region
\begin{eqnarray}
\lefteqn{B_5(\alpha_1,\alpha_2,\alpha_3,\alpha_4,\alpha_5)  = } \nonumber\\
&=&\!\! \iint\limits_{0<y<x<1} x^{\alpha_1-\alpha_2-\alpha_5}
   y^{\alpha_2-1}(1-x)^{\alpha_3-1}
   (x-y)^{\alpha_5-1}(1-y)^{\alpha_4-\alpha_3-\alpha_5}\,dx\,dy \,
\label{triB5.eq}
\end{eqnarray}
for suitably restricted values of the arguments
$(\alpha_1,\alpha_2,\alpha_3,\alpha_4,\alpha_5)$. 
The discovery of this function
indicated that the Euler Beta function was not alone: the Euler Beta function,
which would now be written as $B_{4}(\alpha_{1},\alpha_{2})$, was henceforth
to be regarded as  the first member of the family of $N$-point functions
$B_{N}$ that might be expected to have interesting properties in
analysis as well as in the quantum theory of relativistic
elementary particles.

\paragraph{Cross-Ratio Coordinates.}
A very rapid series of steps subsequently led to what became the
standard Koba-Nielsen representation \cite{KobaNielsen1969b} for the
$N$-point function $B_{N} (\alpha_{13}, \ldots, \alpha_{N-2,N})$,
which can be written as an $(N-3)$-dimensional integral
\begin{equation}
B_{N} (\alpha_{13}, \ldots, \alpha_{N-2,N})
 =  \idotsint\limits_{0<t_1<\cdots<t_{N-3}<1}
 \prod_{i,j} u_{ij}^{\;\;\;\alpha_{ij}-1} d\mu_{N} \ ,
\label{BNintegral.eq}
\end{equation}
where $d\mu_{N} = {dt_{1} \cdots d t_{N-3}}/
         {\prod_{i=1}^{N-4} t_{i} \prod_{j=2}^{N-3} (1 - t_{j})}$
and the $u_{ij}$ are the $N$-point cross-ratios parameterized by
$t_1,\ldots,t_{N-3}$ as described in detail in Section
 \ref{sec:crossratios}.  The formulas (\ref{B4.eq}) and
(\ref{triB5.eq}) correspond to (\ref{BNintegral.eq}) for the cases
         $N=4$ and $N=5$, respectively.


A variety of methods have been employed to study the properties of the
$B_{N}$ integrands as functions of complex variables.  For example,
Koba and Nielson \cite{KobaNielsen1969b} expressed
(\ref{BNintegral.eq}) as an integral  in a space
that was essentially a product of $(N-3)$ copies of $\CP{1}$.  As
noted by one of the current authors in \cite{hansonPGL72}, one can
alternatively express the complex integrand by employing
$\CP{N-3}$ cross-ratios (with a much larger symmetry group) in place of
the product of $(N-3)$ complex projective lines with the single shared
linear fractional transformation symmetry characterizing the
Koba-Nielsen framework.

 We will see in the following that, for $B_{5}$, the compactification
of the set of all five-point cross-ratios can be identified with
$\RP{2}\# 4{\RP{2}}$ as an algebraic subvariety in $\RP{5}$ with a
polyhedral structure that has 12 pentagonal faces.  This embedding of
$\RP{2}\# 4{\RP{2}}$ has a symmetry group of order 120 in $\PGL{6}$.
The double covering, which is the surface of genus four embedded in
$\Sphere{5}\subset \mathReal^{6}$, has a corresponding polyhedral
structure with 24 pentagonal faces and a symmetry group of order 240
in $\OO{6}$.  The integral (\ref{triB5.eq}) is taken over one of the
12 pentagonal faces of $\RP{2}\# 4{\RP{2}}$.  This is the starting
point for the contour integral representation of $B_{5}$.

The study of such a tessellation on $\RP{2}\# 4{\RP{2}}$, the five
crosscap surface, and its symmetry group dates back to the 19th
century \cite{KleinIkosa} and is treated in detail in the work of
Brahana and Coble in 1926 \cite{Brahana1926}.  It is interesting to
see that our space of five-point cross-ratios leads naturally to the
same tessellation, and to the presentation of the symmetry group in
$\OO{6}$.

\paragraph{Contour Integral Representations.}
It is well known that the analytic continuation of the function
defined by (\ref{B4.eq}) is a meromorphic function of
$(\alpha_{1},\alpha_{2})$  on the entire complex space $\mathComplex^2$.
In fact,  changing variables in the integral allows the Beta function
to be rewritten in terms of the standard integral representation of
the Gamma function, leading to the explicit analytic continuation formula
\begin{equation}
B(\alpha_{1},\alpha_{2}) = \frac{\Gamma(\alpha_{1})
  \Gamma(\alpha_{2})}{\Gamma(\alpha_{1}+\alpha_{2})} \ . 
\label{B4gamma.eq}
\end{equation}

In 1890, Pochhammer \cite{pochhammer1890} gave another interesting
continuation formula for $B(\alpha_{1},\alpha_{2})$ in the following form,
\begin{equation}
B(\alpha_{1},\alpha_{2})= \frac{\epsilon(\alpha_{1},\alpha_{2})}
   {(1-e^{2\pi i \alpha_{1}})(1-e^{2\pi i \alpha_{2}})} \ ,
\label{pochratio.eq}
\end{equation}
where $\epsilon(\alpha_{1},\alpha_{2})$ is a contour integral of
$\beta(z;\alpha_{1},\alpha_{2})=z^{\alpha_{1}-1}(1-z)^{\alpha_{2}-1}$
along a properly immersed loop in $\mathComplex \setminus \{0,1\}$,
and hence is a holomorphic function of $(\alpha_{1},\alpha_{2})$.

Our observation that the $B_{5}$ function can be expressed by an
integral over one pentagonal face of $\RP{2}\# 4{\RP{2}}$
leads to a contour integral representation analogous to Pochhammer's
classic representation of $B_4$.  We obtain the following
two-dimensional contour integral representation of $B_{5}$:
\begin{eqnarray}
\lefteqn{B_{5}(\alpha_1,\alpha_2,\alpha_3,\alpha_4,\alpha_5) = }\nonumber\\
&=& \frac{\epsilon(\alpha_1,\alpha_2,\alpha_3,\alpha_4,\alpha_5)}
{ (1-e^{2\pi i \alpha_1})(1-e^{2\pi i \alpha_2})(1-e^{2\pi i \alpha_3})
(1-e^{2\pi i \alpha_4})(1-e^{2\pi i \alpha_5})} \ .
\label{B5ratio.eq}
\end{eqnarray}
Here $\epsilon(\alpha_1,\alpha_2,\alpha_3,\alpha_4,\alpha_5)$ is a holomorphic function
expressed as an integral of a holomorphic 2-form along a closed
oriented surface of genus 5 properly immersed in $\CP{2} \# 4\overline{\CP{2}}$.
Note that, unlike the representation (\ref{triB5.eq}), where
$(\alpha_1, \alpha_2, \alpha_3, \alpha_4, \alpha_5)$ must be properly
restricted for the integral 
to be convergent, the representation (\ref{B5ratio.eq}) of $B_{5}$ is
a meromorphic function of
$(\alpha_1,\alpha_2,\alpha_3,\alpha_4,\alpha_5)$ and is defined on 
the entire space $\mathComplex^5$. Hence the formula
(\ref{B5ratio.eq}) is an explicit analytic continuation formula for
the five-point function $B_{5}$ originally defined by
(\ref{triB5.eq}).

We point out that, to produce the required contour for $B_{5}$, not
only is the two-complex-variable environment supplied by the
Koba-Nielson product of two projective lines, $\CP{1} \times \CP{1}$,
inadequate, but the  richer alternative
$\CP{2}$ framework of \cite{hansonPGL72} is {\it also\/} inadequate.
 The contour lies instead in $\CP{2} \# 4\overline{\CP{2}}$,
 which is the complexification of the
above-mentioned five-crosscap surface $\RP{2} \# 4{\RP{2}}$
in $\RP{5}$ considered as the real part of $\CP{5}$.

We begin in Section \ref{sec:crossratios} by introducing the $N$-point
cross-ratio, which gives rise to the subvarieties upon which our
analysis is based.  Section \ref{sec:b5closure} constructs the
12-pentagon tessellation of the five-crosscap surface as the
compactification of the set of 5-point cross-ratios; its symmetries
and the genus-four double cover are given in Section
\ref{sec:symmetries}.  Then, in Section \ref{sec:B4Poch}, we review
Pochhammer's classical construction for the contour integral
representation of the Euler Beta function.  The framework for studying
$B_{5}$ is set up in Section \ref{sec:B5function} where the
representation (\ref{B5ratio.eq}) is proven.  Selected constructions
are applied to visualizations and computer graphics representations of
the relevant structures in Section \ref{sec:Vis}.
Remarks on the extension to general $N$ are presented in Section
\ref{sec:BNfcn}.

\section{Cross-Ratios}
\label{sec:crossratios}

Recall that the cross-ratio of four distinct ordered numbers
 $\{w,x,y,z\}\subset\mathReal\cup\{\infty\}$  is defined as
\begin{equation}
u(w,x,y,z) = {\frac{(w-y)}{(w-z)}}{\Bigg \slash}{\frac{(x-y)}{(x-z)} } =
\frac{(w-y)(x-z)}{(w-z)(x-y)} \ .
\label{crossratiodef.eq}
\end{equation}
 For any integer $N\geq 4$, we define the
{\it $N$-point cross-ratio\/} of a cyclically-ordered set of $N$ distinct numbers
$\{x_1,\ldots,x_{N}\} \subset \mathReal \cup \{\infty\}$  as the
ordered set of $N(N-3)/2$ numbers
$(u_{13}, u_{14}, \ldots, u_{N-2,N} )$, where
\begin{eqnarray}
u_{ij} &= & u(x_{i},x_{i+1},x_{j+1},x_{j}) =
  \frac{(x_{i}-x_{j+1})(x_{i+1}-x_{j})}{(x_{i}-x_{j})(x_{i+1}-x_{j+1})}
\label{crossratioxi.eq}
\end{eqnarray}
and $ 1\leq i < j \leq N$, $2\leq j-i \leq N-2 $.

The set of all $N$-point cross-ratios can be considered as a subset of
$\mathReal^{N(N-3)/2}$, which we denote by $\mathcal{C}$.  From the
well-known fact that the cross-ratio is invariant under linear
fractional transformations of $\mathReal\cup\{\infty\}$, it is clear
that $\mathcal{C}$ can be parameterized by $(N-3)$ variables, which we
denote as $(t_{1},\ldots,t_{N-3})$.  That is, each point of
$\mathcal{C}$ is the $N$-point cross-ratio of the $N$ cyclically
ordered numbers
\begin{equation}
\{0,\infty,1,t_{1},\ldots,t_{N-3}\},
\label{Nvarlist.eq}
\end{equation}
for a unique  $(t_{1},\ldots,t_{N-3})\in \mathReal^{N-3}$, where
$ t_{1},\ldots,t_{N-3}$ are distinct and not equal to 0 or 1.

\begin{figure}[!htb]
\centering
  \mbox{\psfig{width=4in, figure=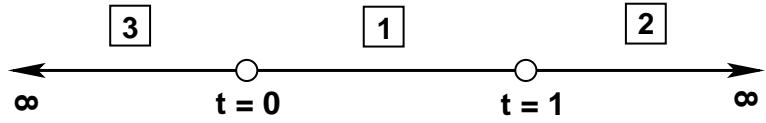}}
  \caption[]{The $3$ connected components of the domain of the
    parameters for the set of 4-point cross-ratios.
   }
  \label{b4-3conn.fig}
\end{figure}

For example, for $N=4$, if we set $x_{1}=0$, $x_{2}=\infty$,
 $x_{3}=1$, and $x_{4}=t$, then, according to (\ref{crossratioxi.eq}),
the set of $4$-point cross-ratios in $\mathReal^{2}$ is given by the
following parameterized curve:
  \begin{eqnarray*}
u_{13} & = & u(x_1,x_2,x_4,x_3) = u(0,\infty,t,1) = t \\
u_{24} & = & u(x_2,x_3,x_1,x_4) = u(\infty,1,0,t) = 1-t  \ .
\end{eqnarray*}
Notice that there are three connected components for the domain of
$t$, as shown in Figure \ref{b4-3conn.fig}.

\begin{figure}[!b]
\centering
  \mbox{\psfig{width=2.8in, figure=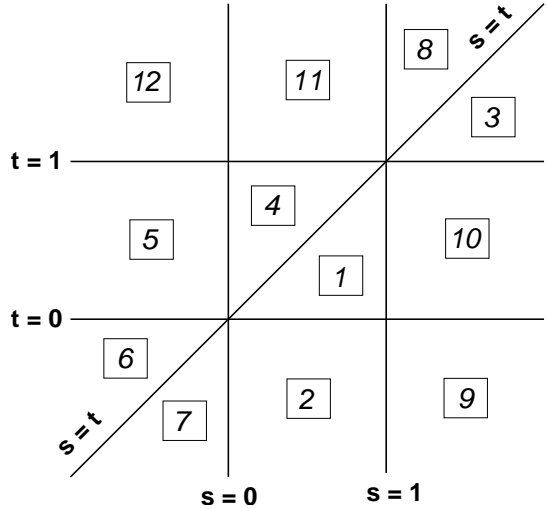}}
  \caption[]{The $12$ connected components of the domain of parameters
 for the set of 5-point cross-ratios.}
  \label{12st.fig}
\end{figure}

For $N=5$, the  set of $5$-point cross-ratios is a surface in
$\mathReal^{5}$  parameterized by $(s,t)$ as
\begin{eqnarray}
u_{13} & = & u(x_1,x_2,x_4,x_3) = u(0,\infty,s,1) = s \nonumber \\
u_{14} & = & u(x_1,x_2,x_5,x_4) = u(0,\infty,t,s) = \frac{t}{s}
\nonumber \\
u_{24} & = & u(x_2,x_3,x_5,x_4) = u(\infty,1,t,s) = \frac{1-s}{1-t} \label{ust5.eq}\\
u_{25} & = & u(x_2,x_3,x_1,x_5) = u(\infty,1,0,t) = 1-t \nonumber \\
u_{35} & = & u(x_3,x_4,x_1,x_5) = u(1,s,0,t) = \frac{s-t}{s(1-t)} \nonumber \ .
\end{eqnarray}
The domain of $(s,t)$ has twelve connected components, as shown in
Figure \ref{12st.fig}.

  One can  verify that the cross-ratios $u_{ij}$
defined by (\ref{crossratioxi.eq}) satisfy
\begin{equation}
u_{ij} = 1 - \prod_{m=i+1}^{j-1} \, \prod _{n=j+1}^{i-1} u_{mn}
\label{uijconstr.eq}
\end{equation}
with the convention that $u_{mn} = u_{nm}$ and $u_{m,n+N} =
u_{m,N}$ for all $1 \leq m,n \leq N$.  In fact, the affine algebraic
subvariety in $\mathReal^{N(N-3)/2}$ defined by (\ref{uijconstr.eq}),
minus a set of measure zero, is  precisely the set
$\mathcal{C}$ of $N$-point cross-ratios.

 In particular, for $N=4$, the constraint (\ref{uijconstr.eq})  becomes
\begin{equation}
 u_{13} = 1 -  u_{24}  \ .
\label{b4uij.eq}
\end{equation}
The set $\mathcal{C}$  is the affine algebraic
subvariety in $\mathReal^{2}$ with coordinates $(z_{1},\,z_{2})$ given
by the linear equation
\begin{equation}
1 - z_1 - z_2  = 0 \ .
\label{b4zi.eq}
\end{equation}

For $N=5$, we have
\begin{eqnarray}
u_{13} & = & 1 - u_{24} \, u_{25}  \nonumber\\
u_{14} & = & 1 - u_{25} \, u_{35} \nonumber\\
u_{24} & = & 1 - u_{35} \, u_{13} \label{5uij.eq}\\
u_{25} & = & 1 - u_{13} \, u_{14} \nonumber\\
u_{35} & = & 1 - u_{14} \, u_{24} \nonumber \ ,
\end{eqnarray}
and $\mathcal{C}$ is the affine algebraic subvariety in
$\mathReal^{5}$ with coordinates
$(z_{1},z_{2},z_{3},z_{4},z_{5})$ given by
\begin{eqnarray}
1 - z_1  - z_3 z_4  & = & 0 \nonumber \\
1 - z_2  - z_4 z_5  & = & 0 \nonumber \\
1 - z_3  - z_5 z_1  & = & 0 \label{b5zi.eq} \\
1 - z_4  - z_1 z_2  & = & 0 \nonumber \\
1 - z_5  - z_2 z_3  & = & 0 \nonumber \ .
\end{eqnarray}

{\it Remark.\/} It can be verified that the system (\ref{b5zi.eq})
has rank 3 at the zero locus, and therefore does actually define a
smooth algebraic subvariety of dimension 2.
\comment{ See file DoubleB5.nb; compute Jacobian, solve for zero
  locus, compute minors, determine rank.}

  Now consider the corresponding projective subvarieties.  For $N=4$,
(\ref{b4zi.eq}) becomes
\begin{equation}
z_{0} - z_1 - z_2   = 0 \ ,
\label{b4projzi.eq}
\end{equation}
which  obviously defines a projective line in $\RP{2}$
with homogeneous coordinates $[z_{0},z_{1},z_{2}]$.

\begin{figure}[!bh]
\centering
  \mbox{\psfig{width=3.0in, figure=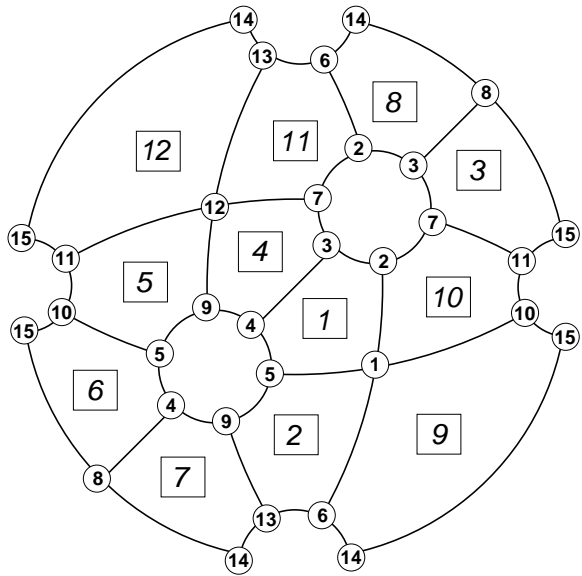}}
  \caption[]{The pentagonal tessellation of the closure of the set of
 5-point cross-ratios.}
  \label{b5-12reg.fig}
\end{figure}

Similarly, for  $N=5$, (\ref{b5zi.eq}) yields
the following homogeneous quadratic  equations
in the homogeneous coordinates $[z_0,z_1,z_2,z_3,z_4,z_5]$ of $\RP{5}$:
\begin{eqnarray}
  z_{0}^{\ 2} - z_0 z_1 - z_3 z_4 & = & 0 \nonumber\\
  z_{0}^{\ 2} - z_0 z_2 - z_4 z_5 & = & 0 \nonumber\\
  z_{0}^{\ 2} - z_0 z_3 - z_5 z_1 & = & 0 \label{b5projzi.eq} \\
  z_{0}^{\ 2} - z_0 z_4 - z_1 z_2 & = & 0 \nonumber \\
  z_{0}^{\ 2} - z_0 z_5 - z_2 z_3  & = & 0 \nonumber\ .
\end{eqnarray}
One can verify that (\ref{b5projzi.eq}) defines a smooth
two-dimensional subvariety in $\RP{5}$, which we will denote by $M$.
To see the topology of $M$, we examine the parameterization
(\ref{ust5.eq}).  As we will show in detail in Section
\ref{sec:b5closure}, the image of each 
of the twelve connected components of the parameter domain has a
smooth pentagonal closure tessellating $M$ as shown in Figure
\ref{b5-12reg.fig}.
Extending (\ref{b5projzi.eq}) to complex variables
defines a complex algebraic variety  $M^{c}$ that is obviously the
complexification of the real manifold $M$.
$M^{c}$ is $\CP{2}$ with four points blown up
and is topologically homeomorphic to $\CP{2} \# 4\overline{\CP{2}}$.

  The tessellation represented in Figure \ref{b5-12reg.fig} has 12
pentagonal faces, $(12\times 5)/2 = 30$ edges, and $(12\times 5)/4 =
15$ vertices; the Euler number of $M$ is thus $\chi = 15-30+12 = -3$,
and therefore $M$ is the connected sum of five $\RP{2}$'s, i.e., a
sphere with five crosscaps.  Therefore, viewing the five-crosscap
surface as the set of cross-ratios yields a natural tessellation with
12 pentagonal faces, which we can call a ``dodecahedron'' even though
it does not bound a 3-ball.  This tessellation was already described
in detail from the point of view of combinatorial topology in the 19th
century \cite{KleinIkosa}.  In 1926, Brahana and Coble
\cite{Brahana1926}, also arrived at the same tessellation of a sphere
with five crosscaps as a map of 12 countries with five sides, and
studied the symmetry group in detail (see also recent work by Weber
\cite{WeberDodec2005} for additional historical background).  Such
tessellations were generalized by Stasheff for use in his study of the
homotopy theory of H-spaces
\cite{StasheffTAMS1963,StasheffHSpaces,Stasheff2004}, and, in
particular, the analogous tiles in higher dimensions are called
associahedra.  These have played a prominent role, e.g., in the work
of Devadoss \cite{Devadoss1999}.  Our discovery of the relation
between the five-crosscap dodecahedral tessellation and the 5-point
cross-ratios, as well as the apparent relation between the
higher-dimensional analogs and the $N$-point cross-ratios, should thus
be of further interest.

\section{Closure of the 5-Point Cross-Ratio Set in $\RP{5}$}
\label{sec:b5closure}

  We now present a detailed treatment of the pentagonal  tessellation
for $M$.  In the homogeneous coordinates
of $\RP{5}$, we will write the parameterization (\ref{ust5.eq}) as
\begin{eqnarray}
p(s,t) & = & \left[1,s,\frac{t}{s},\frac{1-s}{1 - t}, (1 - t),
        \frac{s - t}{s(1-t)}\right] \nonumber\\
 & = & \left[s(1 - t), s^2 (1 - t), t(1 - t), s(1 - s), s(1 -
  t)^2, s - t\right] \ .
\label{b5hom.eq}
\end{eqnarray}

On the triangular connected component $0<t<s<1$, as $(s,t)\rightarrow
(0,0)$ or $(s,t)\rightarrow (1,1)$, the images do not converge to
a point.  To extend the parameterization to the boundary of the
domain, we will replace the parameters $(s,t)$ as follows.

\begin{figure}[htb]
\parbox[t]{5.0in}{\centering
\mbox{\psfig{width=2.2in, figure=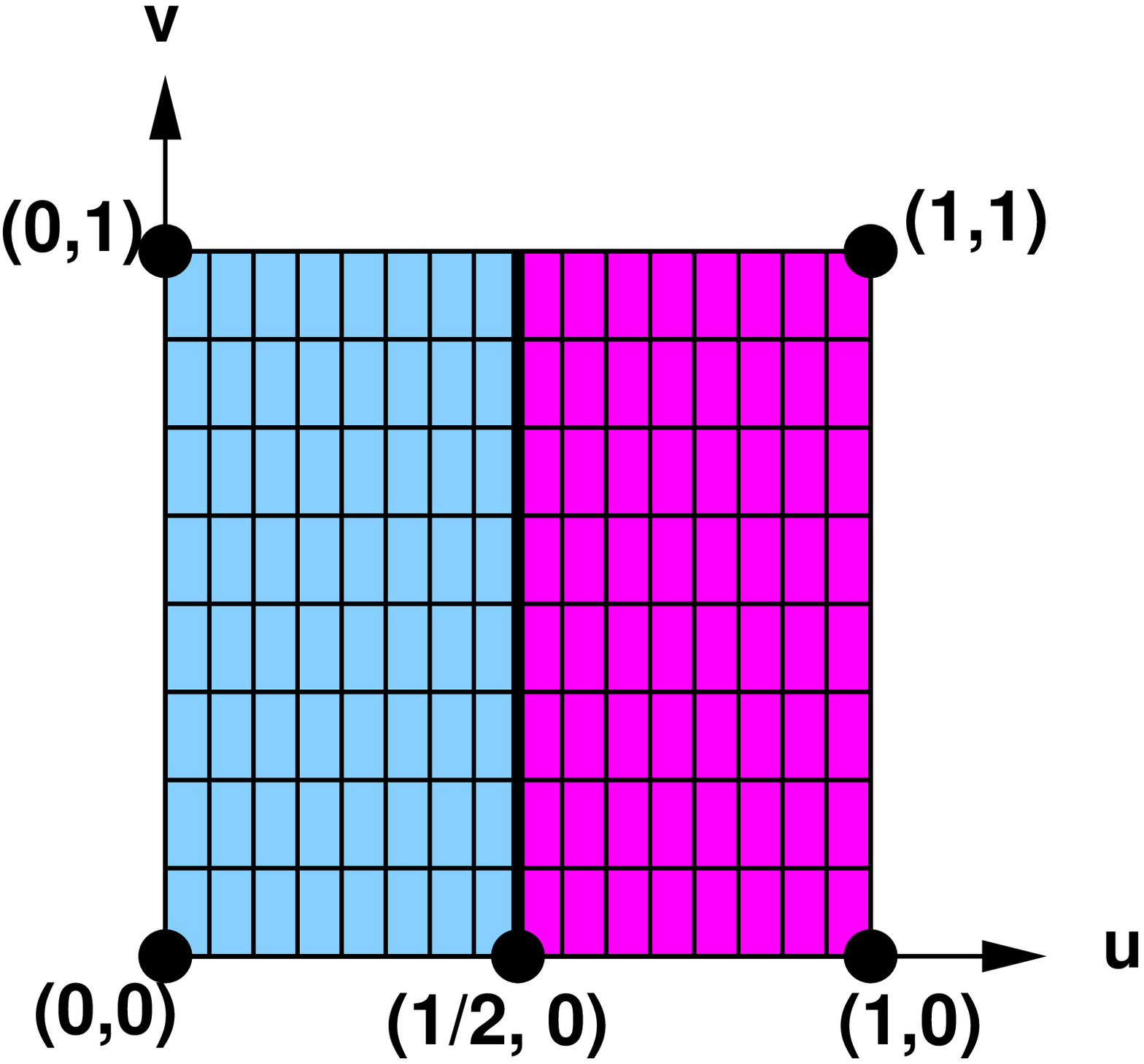}}
\hspace{0.1in}\raisebox{1in}{$\Longrightarrow$}\hspace{0.1in}
\mbox{\psfig{width=2.2in, figure=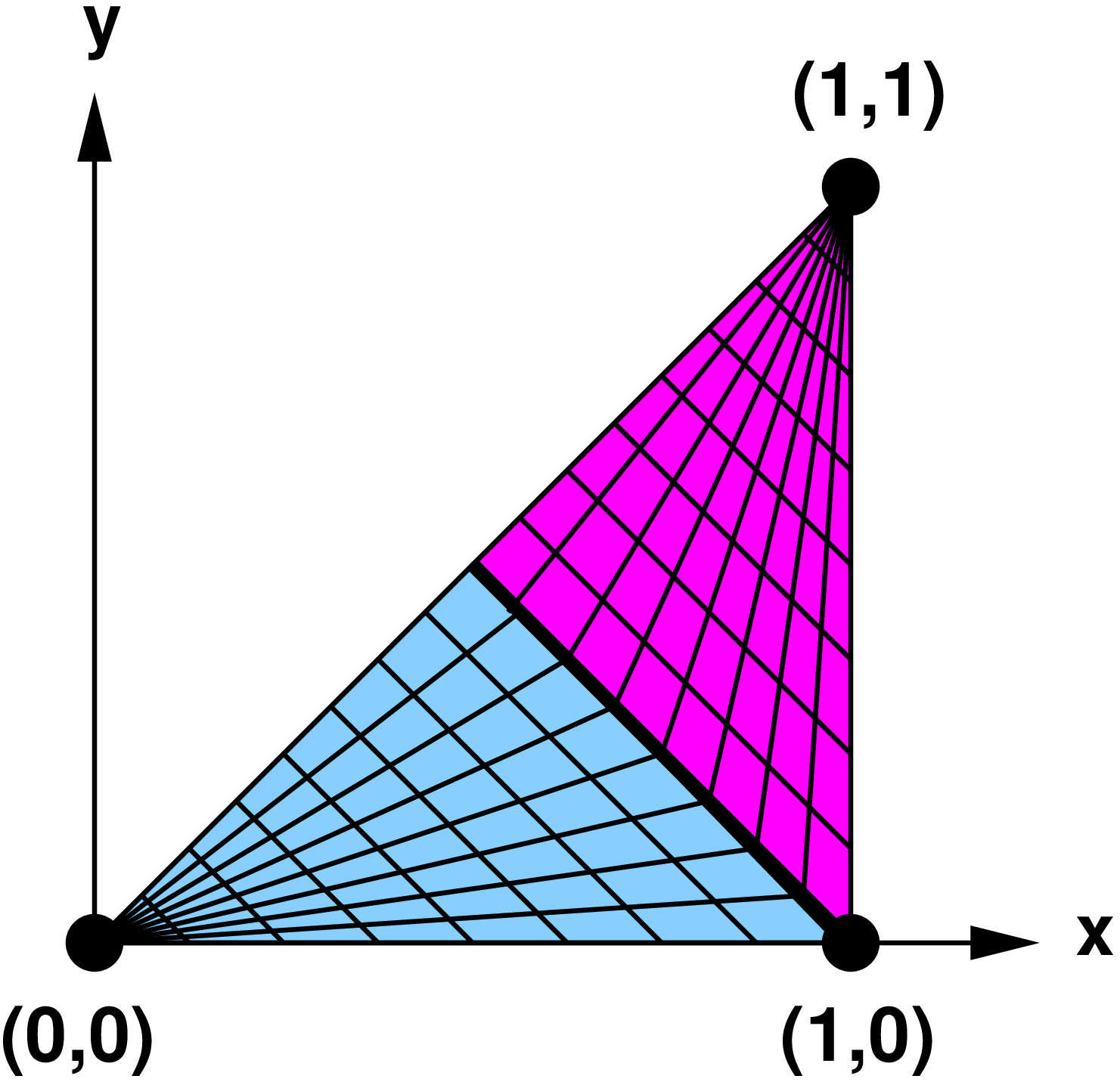}}
  \caption[]{The map from the square $0<u,v<1$  in the $uv$-plane
 to the triangular region $0<y<x<1$ in the $xy$-plane.}
  \label{uvmap.fig}}
\end{figure}

First, let
\begin{equation}
\!\!(x,y) = \left\{\!\!\begin{array}{ll}
                 (2 u - u v,\, u v), & \!\mbox{for $0 < u \leq
                 \frac{1}{2},\,\,0<v<1$ }\\
                (1 - v + u v, \,  -1 + 2 u + v -u v), &
             \!\mbox{for $ \frac{1}{2} \leq u < 1,\,0<v<1$.}
          \end{array} \right.
\label{xyuv.eq}
\end{equation}
The formula (\ref{xyuv.eq}) defines a 1-1 map between the open square
$(0,1)\times (0,1)$ in the $uv$-plane and the open triangle $0<y<x<1$
in the $xy$-plane, as shown in Figure \ref{uvmap.fig}.

Next, in Table \ref{b5-12reg.tbl}, we present twelve formulas that give
1-1 maps between the open
triangular domain $0<y<x<1$ and each of the twelve connected
components shown in Figure \ref{12st.fig}.

\begin{table}[t!]
\begin{centering}
\begin{tabular}{|r|p{4.0in}|} \hline
\raisebox{-1.0ex}{n} &
The map from the region $0<y<x<1$ to the connected component
 \fbox{$n$} in Figure \ref{12st.fig}. \\ \hline
 \raisebox{-5pt}{\rule{0in}{15pt}}
${1}$  & $s=x$, $t=y$ \\                   
 ${2}$ & $s = (x-y)/(1-y),\, t =  y/(y -1)$  \\        
 ${3}$ & $ s  =  1/(x y),\, t  =  x/y $  \\            
 ${4}$ & $  s  =  y,\, t  =  x $  \\                   
 ${5}$  & $  s  =  y/(y-1),\, t  =  (x-y)/(1-y) $ \\   
 ${6}$ &  $  s  =  x/(x-1),\, t  =  (x-y)/(x -1) $ \\  
 ${7}$ &  $ s  =  (x-y)/(x-1),\, t  =  x/(x-1) $  \\   
 ${8}$ &  $  s  =  x/y,\, t  =  1/(x y)  $  \\         
 ${9}$ & $ s  =  (1-y)/(x-y),\, t  =  y/(y-x) $  \\    
${10}$ &  $ s  =  1/(1-y),\, t  =  (1-x)/(1-y) $  \\   
${11}$ & $ s  =  (1-x)/(1-y),\, t  =  1/(1-y)  $  \\   
${12}$ & $  s  =  y/(y-x),\, t  =  (1-y)/(x-y) $  \\   
\hline
\end{tabular}
\caption[]{Transformations from the triangular region $0<y<x<1$ to the
 12 connected components in Figure \ref{12st.fig}.}
\rule{5.375in}{0.01in}
\label{b5-12reg.tbl}
\end{centering}
\end{table}

Composing  (\ref{b5hom.eq}), the entries in Table \ref{b5-12reg.tbl},
and (\ref{xyuv.eq}), we get a parameterization for each of the 12
connected components of $\mathcal{C}$ on the common domain
$(0,1)\times (0,1)$ in the
$uv$-plane.  We will denote these parameterizations by
$f_{1},\ldots,f_{12}$, respectively.

It can be verified that each of the $f_{1},\ldots,f_{12}$ extends as a 1-1
parameterization to the closed square $[0,1]\times [0,1]$.  Each of
the twelve images is a smooth, closed, pentagonal  surface patch
whose vertices correspond to
\[ (u,v) =
({\textstyle\frac{1}{2}},0),\,(1,0),\,(1,1),\, (0,1),\,(0,0) \ . \]

Examining the pentagons one by one, we find they are joined together
to form the closed surface represented by Figure \ref{b5-12reg.fig}.

For future reference, we list below the homogeneous coordinates of the
15 vertices:
\begin{equation}
\begin{array}{rclp{0.5in}rcl}
v_{1} & = & [1, 1,\, 0,\, 0,\, 1,\, 1 ]&&
  v_{9} & = & [0,\, 0,\, 1,\, 0,\, 0,\, -1 ]\\
v_{2} & = & [1,  1,\, 1,\, 0,\, 0,\, 1 ]&&
  v_{10} & = & [0,\, -1,\, 0,\, 1,\, 0,\, 0 ]\\
v_{3} & = & [1, \, 1,\, 1,\, 1,\, 0,\, 0]&&
  v_{11} & = & [0,\, 0,\, 0,\, 1,\, 0,\, 0]\\
v_{4} & = & [1,\, 0,\, 1,\, 1,\, 1,\, 0 ]&&
  v_{12} & = & [0,\, 0,\, 0,\, 0,\, 0,\, 1]\\
v_{5} & = & [ 1, \,0,\, 0,\, 1,\, 1,\, 1 ]&&
    v_{13} & = & [0,\, 0,\, 1,\, 0,\, 0,\, 0 ]\\
v_{6} & = & [0,\, 0,\, -1,\, 0,\, 1,\, 0 ]&&
    v_{14} & = & [0,\, 0,\, 0,\, 0,\, 1,\, 0]\\
v_{7} & = & [0,\, 0,\, 0,\, -1,\, 0,\, 1]&&
  v_{15} & = & [0,\, 1,\, 0,\, 0,\, 0,\, 0] \ .\\
  v_{8} & = & [0, 1,\, 0,\, 0,\, -1,\, 0 ] && &&
\end{array}
\label{origcoords.eq}
\end{equation}

\section{Symmetries and the Double-Covering Lift}
\label{sec:symmetries}

One can see using combinatorial arguments that the tessellation of $M$
shown in Figure \ref{b5-12reg.fig} has many symmetries.  We will
present the group of symmetries as follows.

One of the symmetries, when restricted to face \fbox{$1$}, is a
rotation that transforms the vertices 1,2,3,4,5 to 2,3,4,5,1.  This
symmetry also transforms vertex 12 to vertex 15.  Using the
coordinates of these vertices from  (\ref{origcoords.eq}), we
construct the matrix
\[X_{5} = B \cdot A^{-1}\]
with $A$  the $6\times 6$ matrix
given by $A=[v_1 \, v_2 \, v_3 \, v_4 \,  v_5 \, v_{12}]$, where
$v_{1}$, etc., are written as column vectors whose components are
specified by (\ref{origcoords.eq}).
Similarly, $B=[ v_2 \, v_3 \, v_4 \, v_5 \,  v_1 \, v_{15}]$,
yielding
\begin{equation}
X_5 = \left[\begin{array}{cccccc}
1 &  0 &  0 &  0 &  0 &  0\\
0 &  0 &  1 &  0 &  0 &  0\\
0 &  0 &  0 &  1 &  0 &  0\\
0 &  0 &  0 &  0 &  1 &  0\\
0 &  0 &  0 &  0 &  0 &  1\\
0 &  1 &  0 &  0 &  0 &  0
\end{array} \right] \ ,
\label{permutationsym.eq}
\end{equation}

Another symmetry is the reflection along the edge joining $v_3$ and
$v_4$, which transforms vertices 1,2,3,4,5,12 to 12,7,3,4,9,1.  As
above, one constructs
\begin{equation}
X_2= \left[\begin{array}{cccccc}
 1 & 0 &  0 & 0  & 0 &  -1 \\
 2 & 0 &  -1 & 0  & -1 &  -1 \\
  0 & 0 &  0 & 1 & 0 &  0 \\
  0 & 0 & 1 & 0  & 0 &  0 \\
 2 & -1 &  0 & -1  & 0 & - 1 \\
  0 & 0 &  0 & 0  & 0 &  -1
\end{array} \right]\ .
\label{reflect2sym.eq}
\end{equation}

The following observations are essential: Viewing $X_5$ and $X_2$ as
elements of $\PGL{6}$, one can verify that the zero-locus of
(\ref{b5projzi.eq}) is invariant under the corresponding
transformations of $\RP{5}$.  $X_5$ and $X_2$ generate a group of
order 120, which is isomorphic to the group of automorphisms of $M$
mentioned above.  In other words, we have embedded the automorphism
group of $M$ in $\PGL{6}$.

In fact, $X_5$ and $X_2$ generate a subgroup $G$ of order 120 in
$\GL{6}$.  We let
\begin{equation}
Q = \frac{1}{70} \sum_{g\in G} (g)^{t} \cdot g \ .
\label{invargrpmetric.eq}
\end{equation}
More explicitly,
\begin{equation}
Q = 
 \left[\begin{array}{cccccc}
 20 &  -6 & -6 & -6 & -6 & -6 \\
 -6 &   4 &  1 &  2 &  2 &  1 \\
 -6 &   1 &  4 &  1 &  2 &  2 \\
 -6 &   2 &  1 &  4 &  1 &  2 \\
 -6 &   2 &  2 &  1 &  4 &  1 \\
 -6 &   1 &  2 &  2 &  1 &  4
\end{array} \right] \ .
\label{qtransposeq.eq}
\end{equation}
As is well-known, $Q$ defines a $G$-invariant quadratic form on
$\mathReal^{6}$.  Then the algebraic subvariety in $\mathReal^{6}$ defined by
\begin{eqnarray}
  z_{0}^{\ 2} - z_0 z_1 - z_3 z_4 & = & 0 \nonumber\\
  z_{0}^{\ 2} - z_0 z_2 - z_4 z_5 & = & 0 \nonumber\\
  z_{0}^{\ 2} - z_0 z_3 - z_5 z_1 & = & 0 \label{b5projall.eq} \\
  z_{0}^{\ 2} - z_0 z_4 - z_1 z_2 & = & 0 \nonumber \\
  z_{0}^{\ 2} - z_0 z_5 - z_2 z_3  & = & 0 \nonumber\\
  \left[z_{0}\, z_{1}\, z_{2}\, z_{3}\, z_{4}\, z_{5}\right]\cdot Q \cdot
    \left[z_{0}\, z_{1}\,  z_{2}\, z_{3}\, z_{4}\, z_{5}\right]^{t} & = & 1 \nonumber
\end{eqnarray}
is $G$-invariant and we denote it by $\widetilde{M}$.

Comparing to (\ref{b5projzi.eq}), we see that
$\widetilde{M}$ is  the double covering of $M$ lifted from
$\RP{5}$ to $\mathReal^{6}$ and is therefore topologically the
orientable surface of {\it genus four\/}.  Notice that $\widetilde{M}$ is also
invariant under the action of $-I_{6}$, where $I_{6}$ denotes the
$6\times 6$ identity matrix.  Hence,  $\widetilde{M}$ is invariant under
the group $\widetilde{G}$  generated by $X_{5}$, $X_{2}$, and $-I_{6}$, which has 240
elements in $\GL{6}$.

Let $P$ be a $6 \times 6$ matrix satisfying
\begin{equation}
P^{t} P = Q \ .
\label{PtPeqQ.eq}
\end{equation}
Then $P\widetilde{M} \subset \Sphere{5}$, where
 $\Sphere{5} = \{(x_{0},\ldots,x_{5})\in \mathReal^{6} \, :\,
 x_{0}^{\ 2} + \cdots +  x_{5}^{\ 2} = 1\}$ is the unit sphere in
 $\mathReal^{6}$, and it is invariant under the subgroup of order 240
 in $\OO{6}$ generated by $P X_{5} P^{-1}$,  $P X_{2} P^{-1}$,
and  $-I_{6}$.

With the parameterization $f_{1}(u,v),\ldots,f_{12}(u,v)$ for $M$ from
Section \ref{sec:b5closure}, we can now easily write down the
following parameterization for $\widetilde{M}$:
\begin{equation}
\tilde{f}_{i}^{\pm} = \frac{\pm f_{i}}{\sqrt{f^{t}_{i} \cdot Q \cdot
    f_{i}}} \ \ \ \ i=1,\ldots,12 \ .
\label{r6lift.eq}
\end{equation}
Each $\tilde{f}_{i}^{\pm}$ maps $[0,1]\times [0,1]$ in the $uv$-plane
to a pentagonal surface patch.  This yields a tessellation of
 $\widetilde{M}$ with 24 pentagonal faces.  As mentioned at the end of Section
\ref{sec:crossratios}, such a tessellation for the genus-four
surface has long been known.  Here we have performed this
tessellation symmetrically in $\mathReal^{6}$.

The coordinates of the vertices appearing in the tessellation of
$\widetilde{M}$ can be computed from (\ref{r6lift.eq}) at the points
$(u,v)\,=\, (\frac{1}{2},0),\, (1,0),\,(1,1),\,(0,1),\,(0,0)$.  They
are in fact the same as those presented in (\ref{origcoords.eq}),
together with their negatives, viewed now as coordinates in
$\mathReal^{6}$.

\begin{figure}[!htb]
\centering
  \psfig{width=2.5in, figure=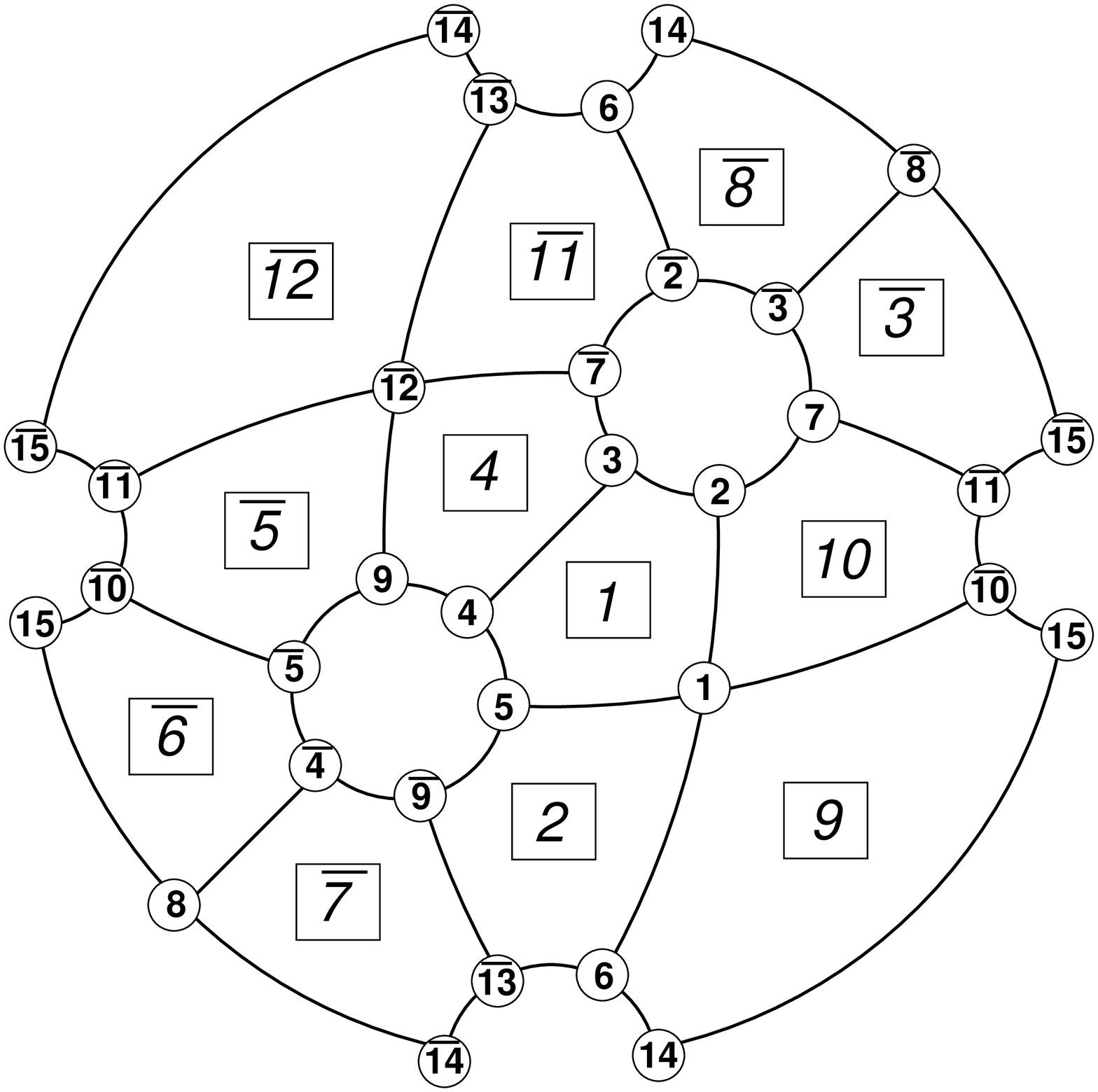}
\hspace{0.1in}
  \psfig{width=2.5in, figure=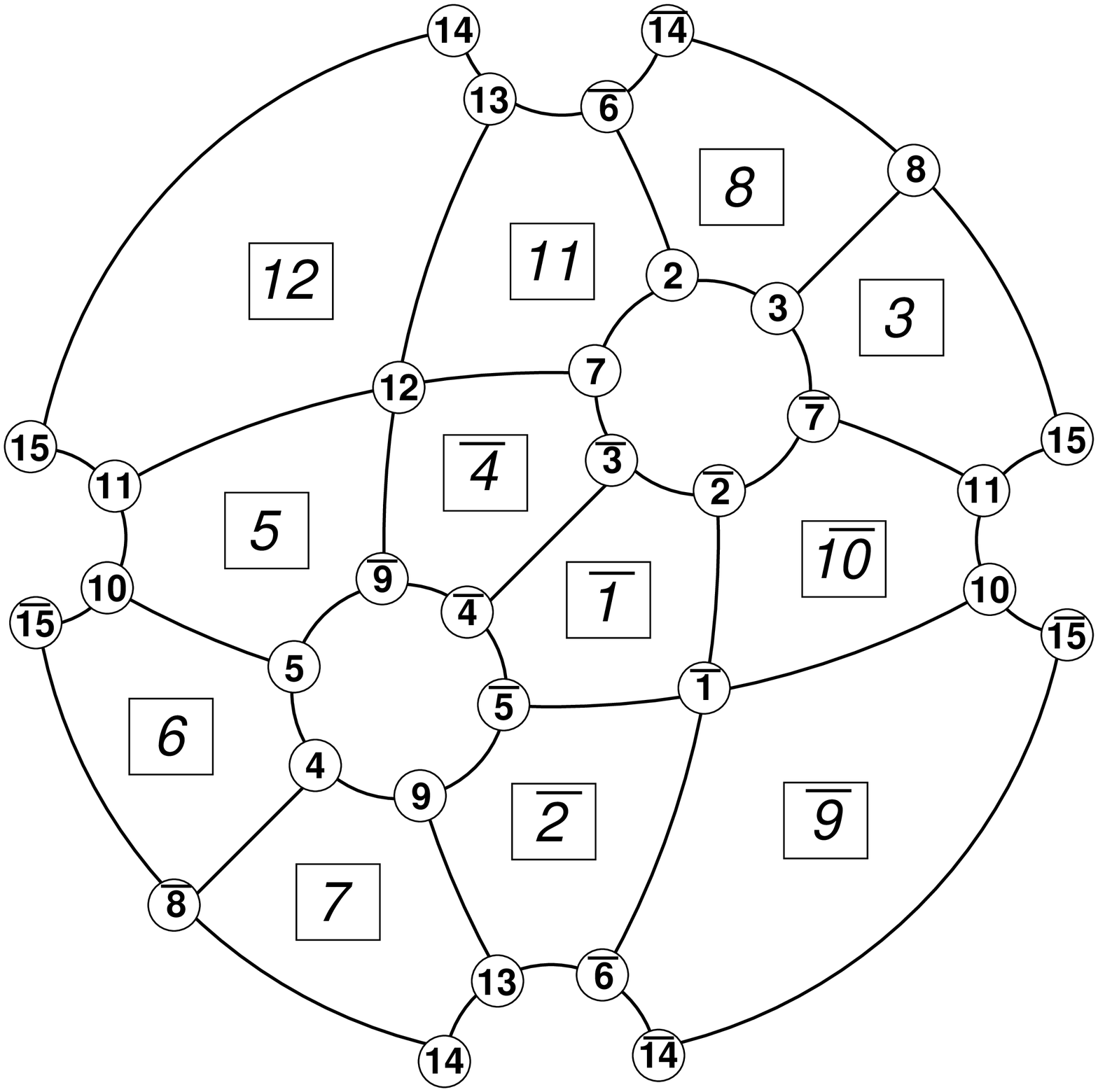}
  \caption[]{The $24$ face identifications of the double covering.}
 \label{b5-12orient.fig}
\end{figure}

We now identify the 24 faces in  $\widetilde{M}$ with ordered sets of
vertices in $\mathReal^{6}$:  the oriented faces are labeled in
terms of the indices of the vertices in (\ref{origcoords.eq}), where a
minus sign indicates the negative mirror
vertex and conjugate  faces are denoted with bars:
\begin{equation}
\begin{array}{rlrl}
\mathrm{face\ } 1: & (1, 2, 3, 4, 5)&
\mathrm{face\ } \bar{1}: & (-1, -5, -4, -3, -2)\\
\mathrm{face\ } 2:& (1,5,- 9, - 13, 6) &
\mathrm{face\ } \bar{2}:& (-1, -6, +13, +9, -5)\\
\mathrm{face\ } 3: & ( 8, 15, 11, -7, 3)&
\mathrm{face\ } \bar{3}: & ( -8, -3, +7, -11, -15)\\
\mathrm{face\ } 4: & (-12, 9, 4, 3, -7) &
\mathrm{face\ } \bar{4}: & (12, 7, -3, -4, -9)\\
\mathrm{face\ } 5: & (12, -9, 5, 10, 11)&
\mathrm{face\ } \bar{5}: & (-12,-11, -10, -5, 9)\\
\mathrm{face\ } 6: & (-8, -15, 10, 5, 4)&
\mathrm{face\ } \bar{6}: & (+8, -4, -5, -10, +15)\\
\mathrm{face\ } 7: &  (-8, 4, 9, 13, 14)&
\mathrm{face\ } \bar{7}: &  (+8,-14, -13, -9, -4)\\
\mathrm{face\ } 8: & (8, 3, 2, -6, -14)&
\mathrm{face\ } \bar{8}: & (-8, +14, +6, -2, -3)\\
\mathrm{face\ } 9:  & (1, 6, 14, 15, -10)&
\mathrm{face\ } \bar{9}:  & (-1, +10, -15, -14, -6)\\
\mathrm{face\ } 10: &  (1, -10, -11, 7, 2)&
\mathrm{face\ } \overline{10}: &  (-1, -2, -7, +11, +10)\\
\mathrm{face\ } 11: & (12, 13, -6, 2, 7)&
\mathrm{face\ } \overline{11}: & (-12, -7, -2, +6,-13)\\
\mathrm{face\ } 12: & (12, 11, 15, 14, 13 ) &
\mathrm{face\ } \overline{12}: & (-12, -13, -14, -15, -11)\ .\\
\end{array}
\label{24facelabels.eq}
\end{equation}
The correspondence between the 12 projective faces and the 12 $(s,t)$
regions of Figure \ref{12st.fig} is shown in Figure  \ref{b5-12reg.fig};
the correspondence between the 24 faces in the double cover
$\widetilde{M}$ and the 12 regions is shown in
Figure \ref{b5-12orient.fig}.

\section{Review of the Pochhammer Contour for B${}_{4}$}\label{sec:B4Poch}

  In this section, we  review Pochhammer's construction
\cite{pochhammer1890,jordan1887,whittakerwatson} of the contour
integral leading to the formula (\ref{pochratio.eq}) for $B_{4}$.
This will lead us to the contour integral representation for $B_{5}$
to be presented in Section \ref{sec:B5function}.

\begin{figure}[!htb]
\centering
  \mbox{\psfig{width=8.5cm, figure=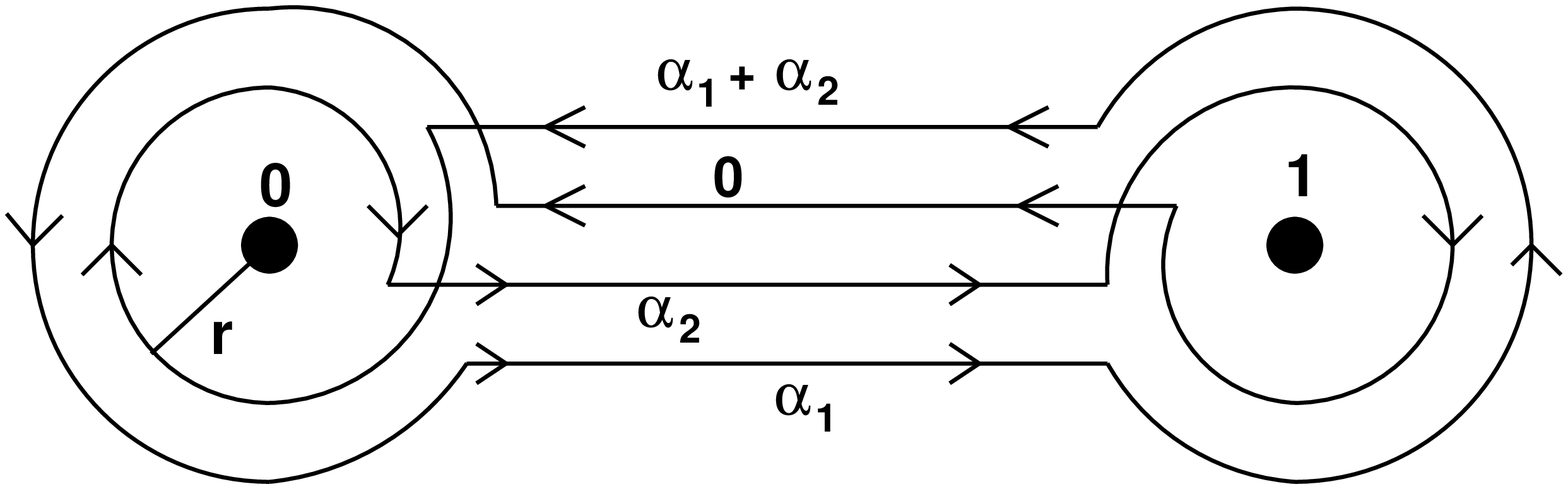}}
  \caption[]{The Pochhammer contour $C$ for the Euler Beta function.}
  \label{pochsketch.fig}
\end{figure}

Let
\begin{equation}
\beta(z;\,\alpha_{1},\alpha_{2}) = z^{\alpha_{1}-1}\,(1-z)^{\alpha_{2}-1} \ .
\label{b4integrand.eq}
\end{equation}
where $(\alpha_{1},\alpha_{2})$ is a pair of arbitrary complex numbers.
 Considered as a function of $z$,
$\beta$ defines a family of holomorphic
functions on a proper Riemann covering sheaf $S$ over
$\mathComplex \setminus\{0,1\}$. Let $C$ be a closed and oriented curve in $S$ that
is the lift of the closed and oriented curve in $\mathComplex
\setminus\{0,1\}$ shown in Figure \ref{pochsketch.fig},  where we label
the line segments by their relative phases in the lift.  This is known as the
 Pochhammer contour.

Next, define
\[\epsilon(\alpha_{1},\alpha_{2})=\int_C\,\beta(z;\,\alpha_{1},\alpha_{2})\,dz\, \ .\]
Clearly $\epsilon(\alpha_{1},\alpha_{2})$ is a holomorphic function of
$(\alpha_{1},\alpha_{2})$ and is 
invariant under continuous deformations of $C$. Therefore,  letting $r
\rightarrow 0^+$ in Figure \ref{pochsketch.fig}, one easily sees that, if
${\rm Re}\,\alpha_{1} >0$ and ${\rm Re}\,\alpha_{2} >0$, then
\[\epsilon(\alpha_{1},\alpha_{2})=(1-e^{2\pi i \alpha_{1}})(1-e^{2\pi
  i \alpha_{2}})\int_0^1\,x^{\alpha_{1}-1}(1-x)^{\alpha_{2}-1}\,dx \ ,
\]
which yields the formula  (\ref{pochratio.eq}).

If $\alpha_{1}$ (or $\alpha_{2}$, resp.) is an integer $\ge 1$, then the holomorphic
1-form $\beta(z;\,\alpha_{1},\alpha_{2})\,dz\,$ on $S$ descends to a holomorphic 1-form
on a proper Riemann covering sheaf over $\mathComplex \setminus \{0\}$
(or $\mathComplex \setminus \{1\}$, resp.). Since the curve in Figure
\ref{pochsketch.fig} is contractible in $\mathComplex \setminus \{0\}$
(or $\mathComplex \setminus \{1\}$, resp.), $\epsilon(\alpha_{1},\alpha_{2})=0$.

\begin {figure}[htb!]
\centering
  \mbox{\psfig{width=8.5cm, figure=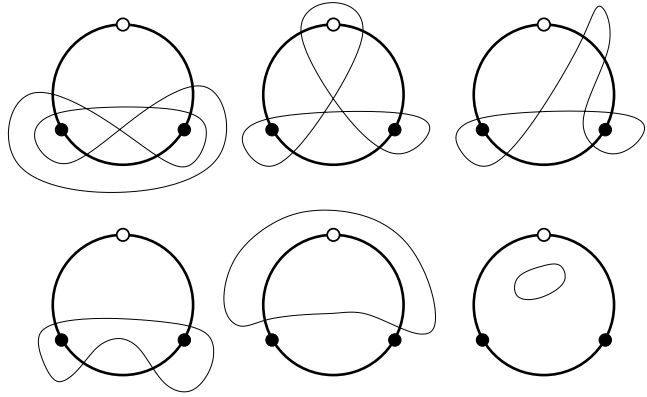}}
  \caption[]{The deformation of the Pochhammer contour to a null
  contour when the conditions $\alpha_{1}+\alpha_{2} = 0,-1,-2,\ldots$ remove
  the branch point at infinity (open circle).}
\label {pochdeform.fig}
\end {figure}

Notice that, by letting  $w=1/z$, we have
\[z^{\alpha_{1}-1}\,(1-z)^{\alpha_{2}-1}\,dz\, =
  -(w-1)^{\alpha_{2}-1}w^{-\alpha_{1}-\alpha_{2}}\,dw\, \ . \]
This shows that, if $(\alpha_{1}+\alpha_{2})$ is a non-positive integer
($\alpha_{1}+\alpha_{2} = 0,\,-1,\,-2,\ldots$),
the holomorphic 1-form $\beta(z;\,\alpha_{1},\alpha_{2})\,dz\,$ on $S$
descends to a holomorphic 1-form on a proper Riemann covering sheaf
over $\CP{1} \setminus\{0,1\}$, where as usual we identify $\CP{1}$
with $\mathComplex\cup\{\infty\}$.  Since the curve in Figure
\ref{pochsketch.fig} is contractible in $\CP{1} \setminus\{0,1\}$
(see Figure \ref{pochdeform.fig}), it therefore follows that
$\epsilon(\alpha_{1},\alpha_{2})=0$ also in this case.   Figure \ref{pochdeform.fig}
shows how the contractibility of the contour can be made explicit.

From these observations and (\ref{pochratio.eq}), one concludes in
particular that the poles of $B(\alpha_{1},\alpha_{2})$ can only occur
at points where either $\alpha_{1}$ or $\alpha_{2}$ is a non-positive
integer. Furthermore, $B(\alpha_{1},\alpha_{2})=0$ if neither
$\alpha_{1}$ nor $\alpha_{2}$ is a non-positive integer, but
$(\alpha_{1}+\alpha_{2})$ is a non-positive integer.  These properties
of course also follow directly from (\ref{B4gamma.eq}); in fact these
are precisely all the poles and zeroes of $B(\alpha_{1},\alpha_{2})$.


\section {Contour Representation of the Function B${}_{5}$}\label{sec:B5function}

We now view $M$ as the real two-dimensional surface in $M^{c}$, as
defined at the end of Section \ref{sec:crossratios}.  The manifold
$M^{c}$ can be visualized by the complexification of Figure
\ref{b5-12reg.fig};  with the edges of the pentagon taken off,
$M^{c}$ is now parameterized by two complex parameters that we denote
as $(z_{1},\,z_{2})$, replacing $(s,t)$ in (\ref{b5hom.eq}).  Following
the procedure in Section \ref{sec:B4Poch}, let
\begin{eqnarray}
\lefteqn{\beta_5(z_1,z_2;\,\alpha_1,\alpha_2,\alpha_3,\alpha_4,\alpha_5)=}\nonumber\\ 
& = &
z_1^{\alpha_1-\alpha_2-\alpha_5}z_2^{\alpha_2-1}(1-z_1)^{\alpha_3-1}
  (z_1-z_2)^{\alpha_5-1}  (1-z_2)^{\alpha_4-\alpha_3-\alpha_5} \ . 
\label{b5integrand.eq}
\end{eqnarray}
Then  (\ref{triB5.eq}) can be viewed as the integral of the (locally)
holomorphic 2-form $\beta_5\,dz_1\wedge dz_2$ (with branched
singularities) on $M^{c}$ over the domain \fbox{$1$} on $M$.

The function $\beta_5$ can be viewed as a
5-complex-parameter family of locally holomorphic functions on $M^{c}$
with branched singularities at the edges of the pentagons evident in the
complexification of Figure \ref{b5-12reg.fig}.

 Let $S$ be the Riemann covering sheaf of $\beta_5$ over $M^{c} \setminus
\{{\rm branch \ lines}\}$. We will construct an orientable closed
surface in $S$ that is the lift of a closed surface in $M^{c}$ obtained
by wrapping properly around the five (complex) edges of the pentagonal
domain \fbox{$1$}.  This will then lead to the formula (\ref{B5ratio.eq}).

\begin {figure}[bh!]
\centering
  \mbox{\psfig{width=5cm, figure=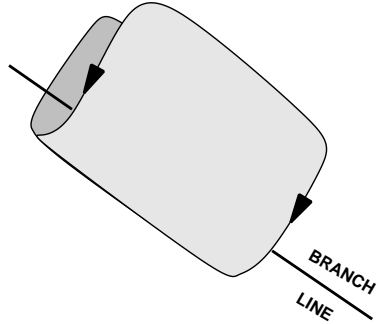}}
\caption [] {A surface sheet folds back around a codimension two branch line.}
\label {foldingbranch.fig}
\end {figure}

A function such as $\beta_5$ is only defined
on $M^{c}$, away from singularities, up to a  factor $e^{2\pi i\gamma}$,
i.e., by a {\it phase} $\gamma$ which is an integer linear combination of
$\alpha_1,\cdots,\alpha_5$.  To lift a surface wrapping around the branch lines to
the covering sheaf $S$, on which $\beta_5$ is a holomorphic function,
we first need to understand how the phase of $\beta_5$ changes on a
piece of surface as it makes a simple fold back around one branch line
(see Figure \ref{foldingbranch.fig}).

It is obvious that if a surface folds back around the branch line
$z_2=0$, $z_1=1$, or $z_1-z_2=0$, then the phase of $\beta_5$
changes by $\pm \alpha_2$, $\pm \alpha_3$, or $\pm \alpha_5$, respectively, where the
sign $+$ or $-$ depends on the folding direction, i.e., whether
the direction is {\it counterclockwise} or {\it clockwise}.

\begin {figure}[ht!]
\centering
  \mbox{\psfig{width=5cm, figure=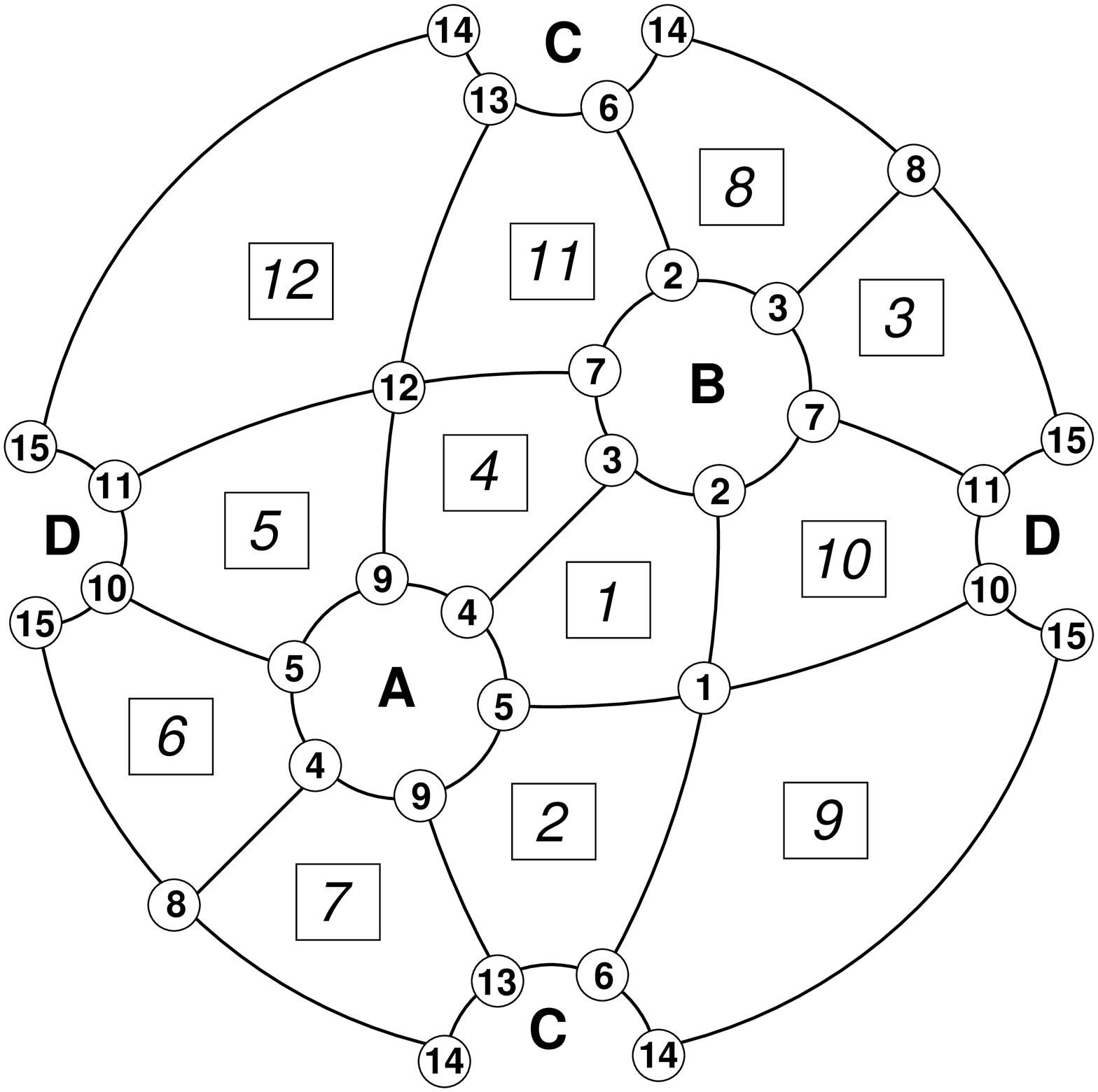}}
\hspace{.5in}
  \mbox{\psfig{width=5cm, figure=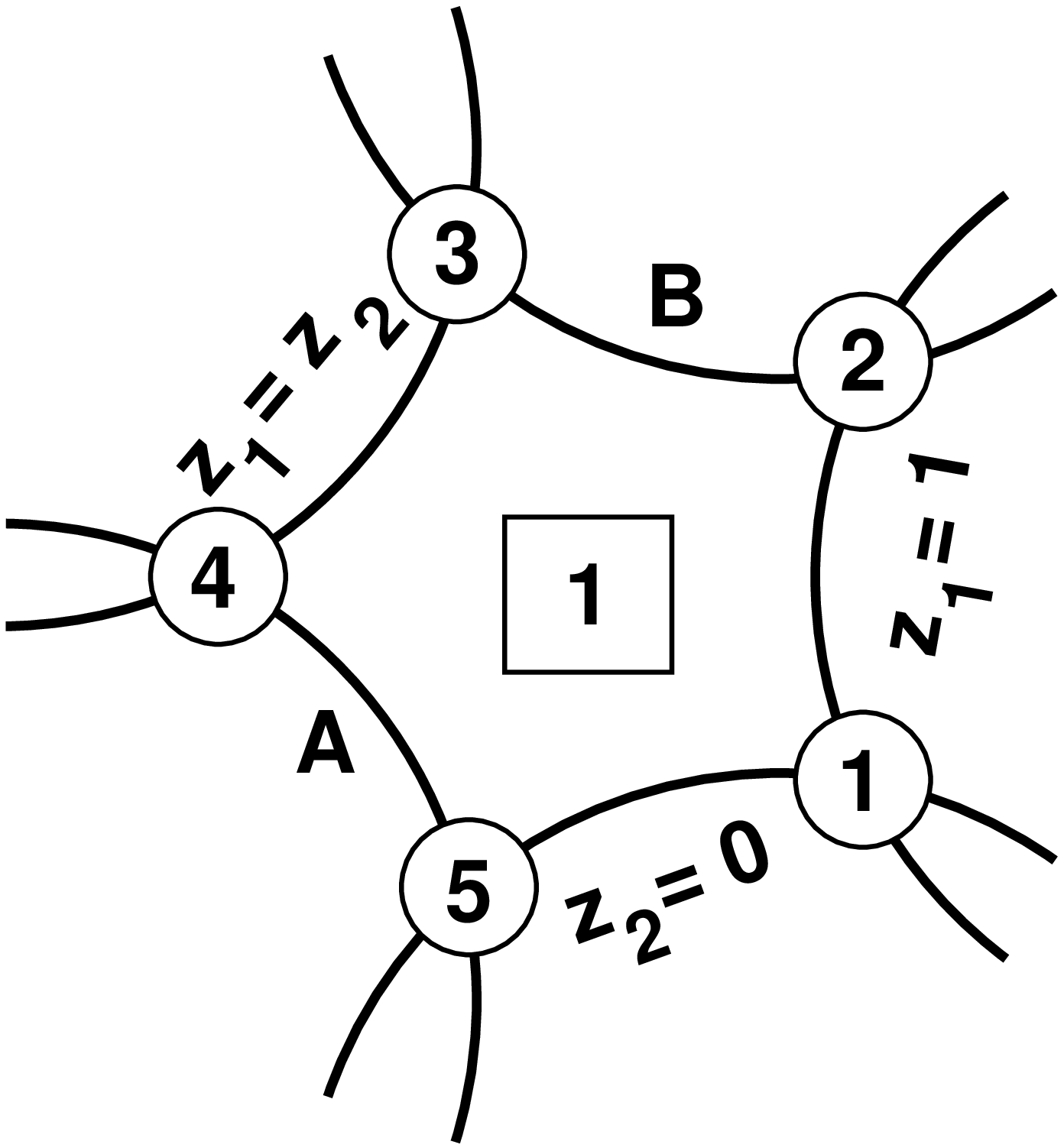}}
\caption [] {The pentagonal domains of the $B_{5}$ integrand, with detailed
pentagonal branch structure shown for region 1.}
\label{cp2blowup.fig}
\end {figure}

\begin {figure}[ht!]
\centering
  \mbox{\psfig{width=4.cm, figure=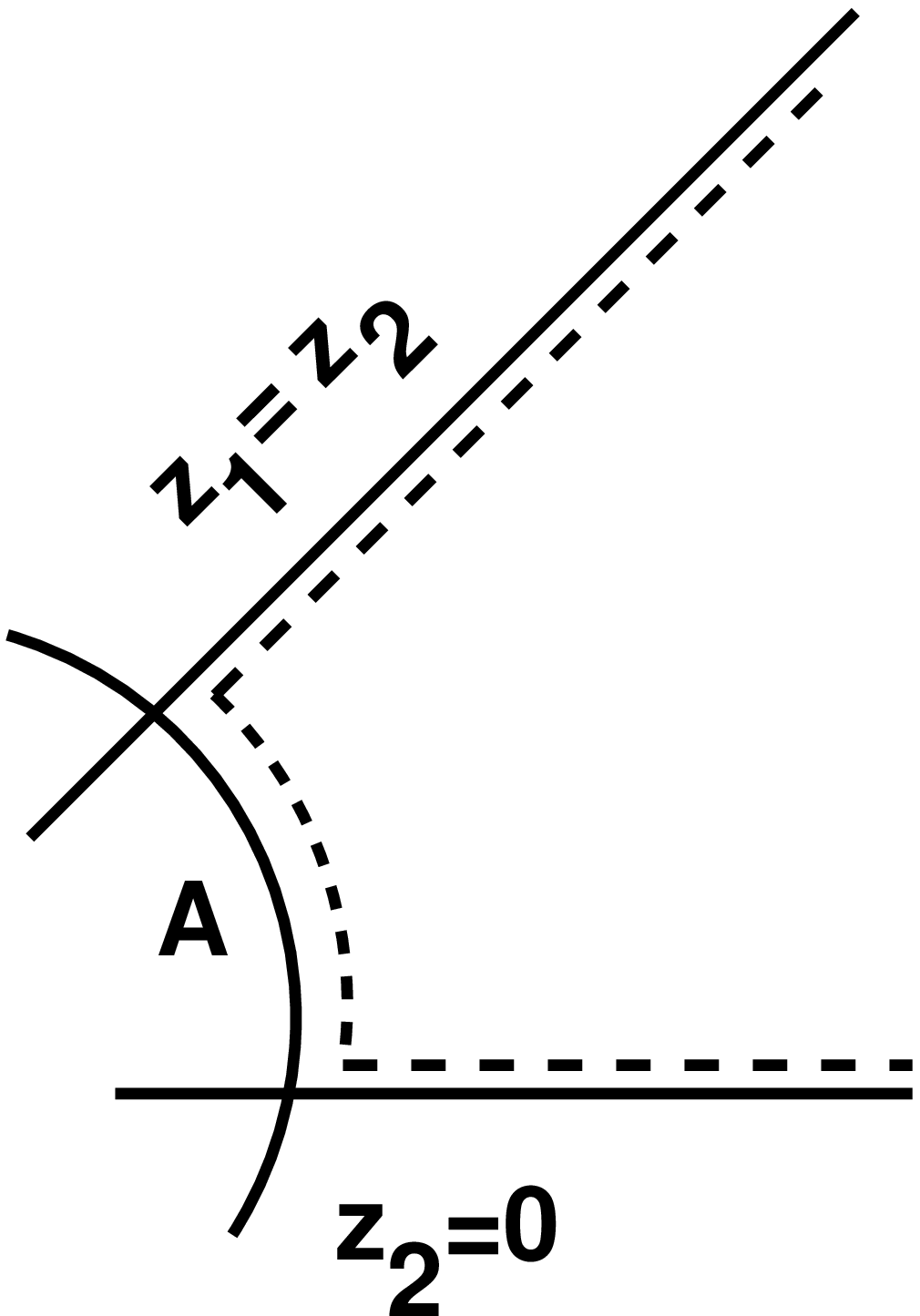}}
  \raisebox{1in}{\hspace{.4in}$\mathbf\Longrightarrow$  \hspace{.4in}}
  \mbox{\psfig{width=5cm, figure=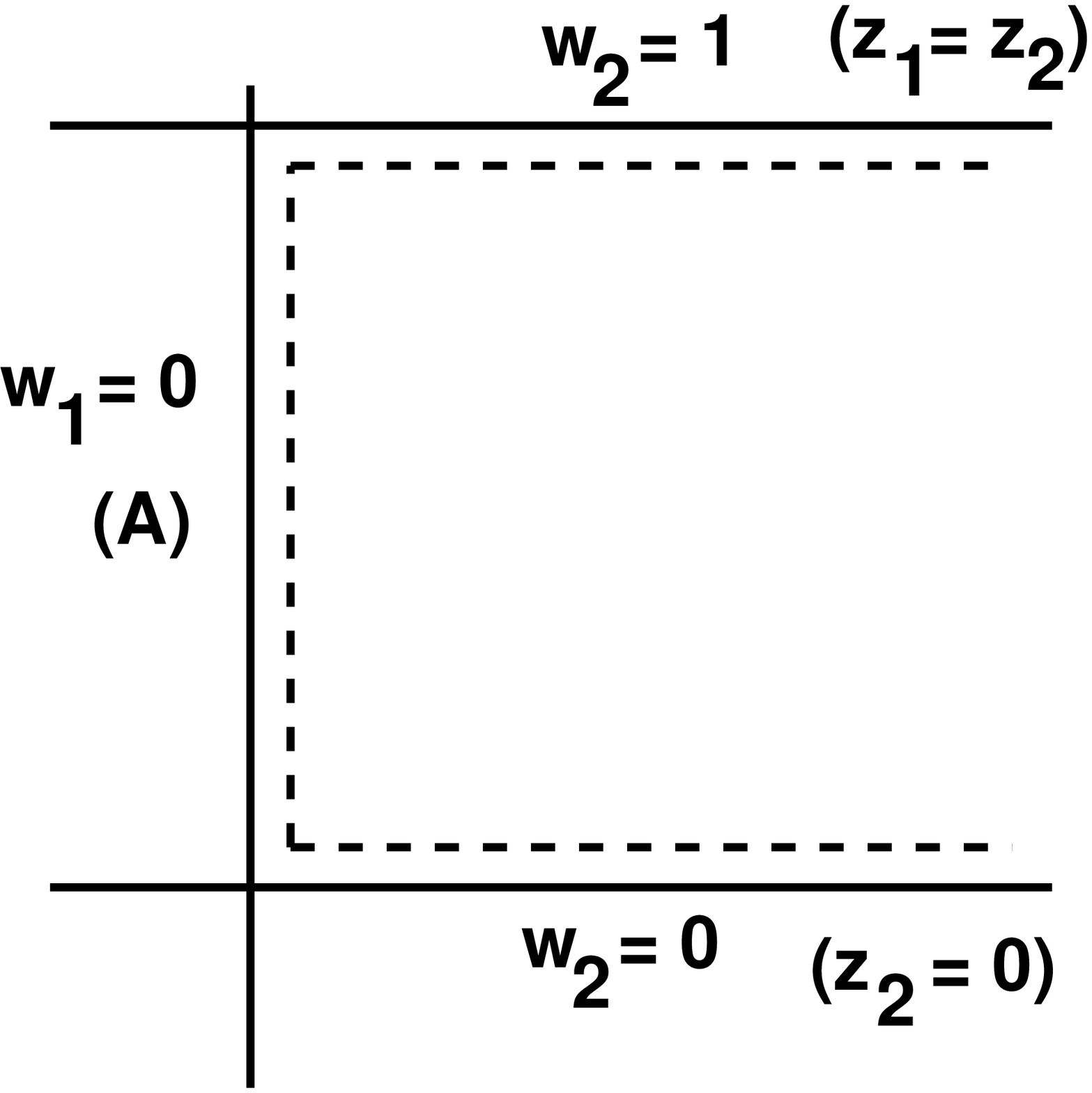}}
\caption [] {The coordinates around the branch line at A.}
\label {blowupA.fig}
\end {figure}

To see how the the phase of $\beta_5$ changes on a surface folding
around the branch line $A$ (see Figure  \ref{cp2blowup.fig}),
let  $(w_1,w_2)$ be the coordinates around
$A$ chosen so that $A$ is given by $w_1=0$, and away from $A$,
$w_1=z_1$, $w_2=z_{2}/z_{1}$  (cf. \cite{GriffithsHarris1978} and
Figure \ref{blowupA.fig}).
 We can then write $\beta_5$ as
\begin{eqnarray*}
\lefteqn{\beta_5(z_1,z_2;\,\alpha_1,\alpha_2,\alpha_3,\alpha_4,\alpha_5)=}\\
& = &
w_1^{\alpha_1-2}w_2^{\alpha_2-1}(1-w_1)^{\alpha_3-1}(1-w_2)^{\alpha_5-1}
(1-w_1w_2)^{\alpha_4-\alpha_3-\alpha_5} \ .
\end{eqnarray*}
It is now easy to see that as a surface
folds back around A, the
phase of $\beta_5$ changes by $\pm \alpha_{1}$.  Similarly, one can show that
as a surface folds back around B, the phase of $\beta_5$ changes by
$\pm \alpha_{4}$.

We now construct an immersed surface in $M$ in three steps as follows:
\begin{itemize}
\item[Step 1.]  Consider a set of 32 copies of the pentagonal sheets
stacked over the region \fbox{1} in Figure \ref{cp2blowup.fig}, with a small
neighborhood of the five corners taken off for now.  From what we have
shown above, it is appropriate to label the edges of each pentagonal
sheet at $A, \ z_2=0, \ z_1=1, \ B, \ z_1-z_2=0$ by $\alpha_1, \ \alpha_2, \
\alpha_3, \ \alpha_4, \ \alpha_5$, respectively (see Figure \ref{pentagonblowup.fig}).

\begin {figure}
\centering
\psfig{width=5cm, figure=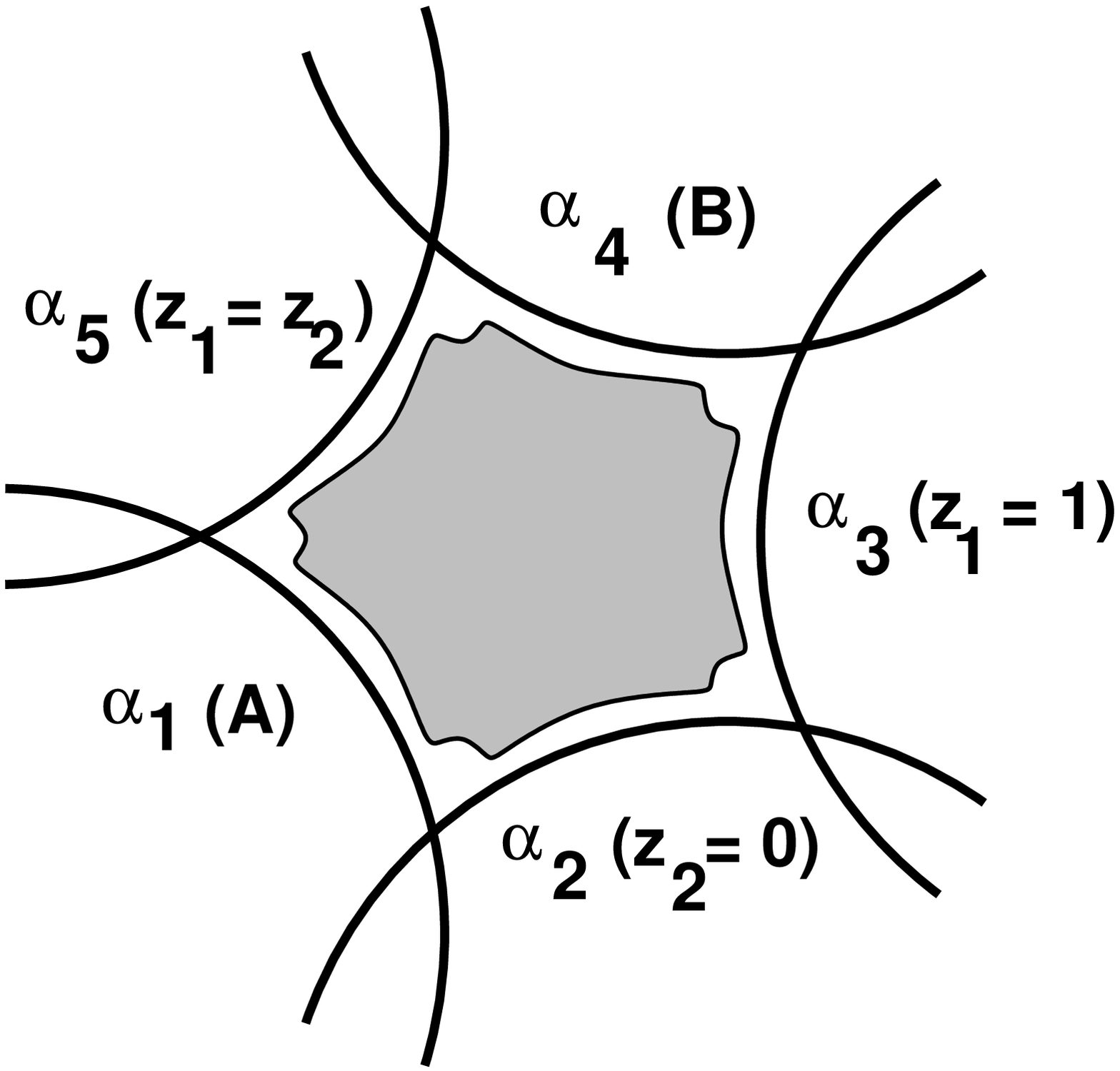}
\hspace{.8cm}
\psfig{width=5cm, figure=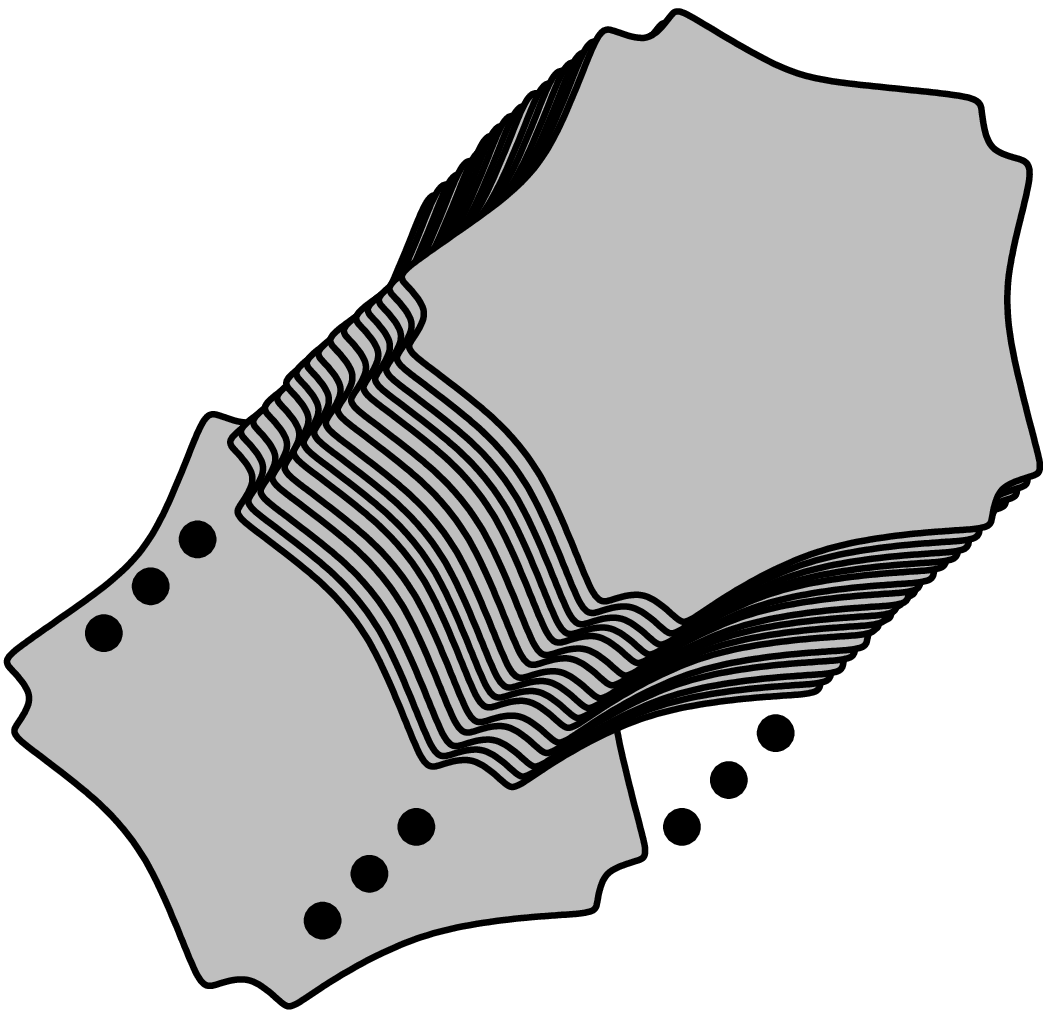}
\caption [] {The pentagonal sheets in  region \fbox{1}.}
\label {pentagonblowup.fig}
\end{figure}

We attach to each of these pentagonal sheets  a {\it phase label}
\[ p_1 \alpha_1+p_2 \alpha_2+p_3 \alpha_3+p_4
\alpha_4+p_5 \alpha_5 \ ,
\]
where $p_j=0$ or $1$ for $j=1,\dots,5$.  Each pentagonal sheet is
given an orientation which is same as or opposite to the original
natural orientation on region \fbox{1} according to whether $\sum p_j$ is
even or odd.

\item[Step 2.]
 Two pentagonal sheets in Step 1 are joined along the
edge $\alpha_j$ by folding around the corresponding branch line in the proper
direction (see Figure \ref{foldingbranch.fig})
if and only if their phase numbers  differ by
$\alpha_j$. It is easy to see that we end up with an immersed oriented
surface in $M \setminus \{{\rm branch \  lines}\}$ that can be lifted
to $S$. However, this surface has  {\it 40 holes\/}
caused by the small neighborhoods that we removed around the corners
where the branch lines intersect.  In Figure \ref{b5commutator.fig},
we show a single instance of one of these holes.

\begin {figure}
\centering
\psfig{width=6cm, figure=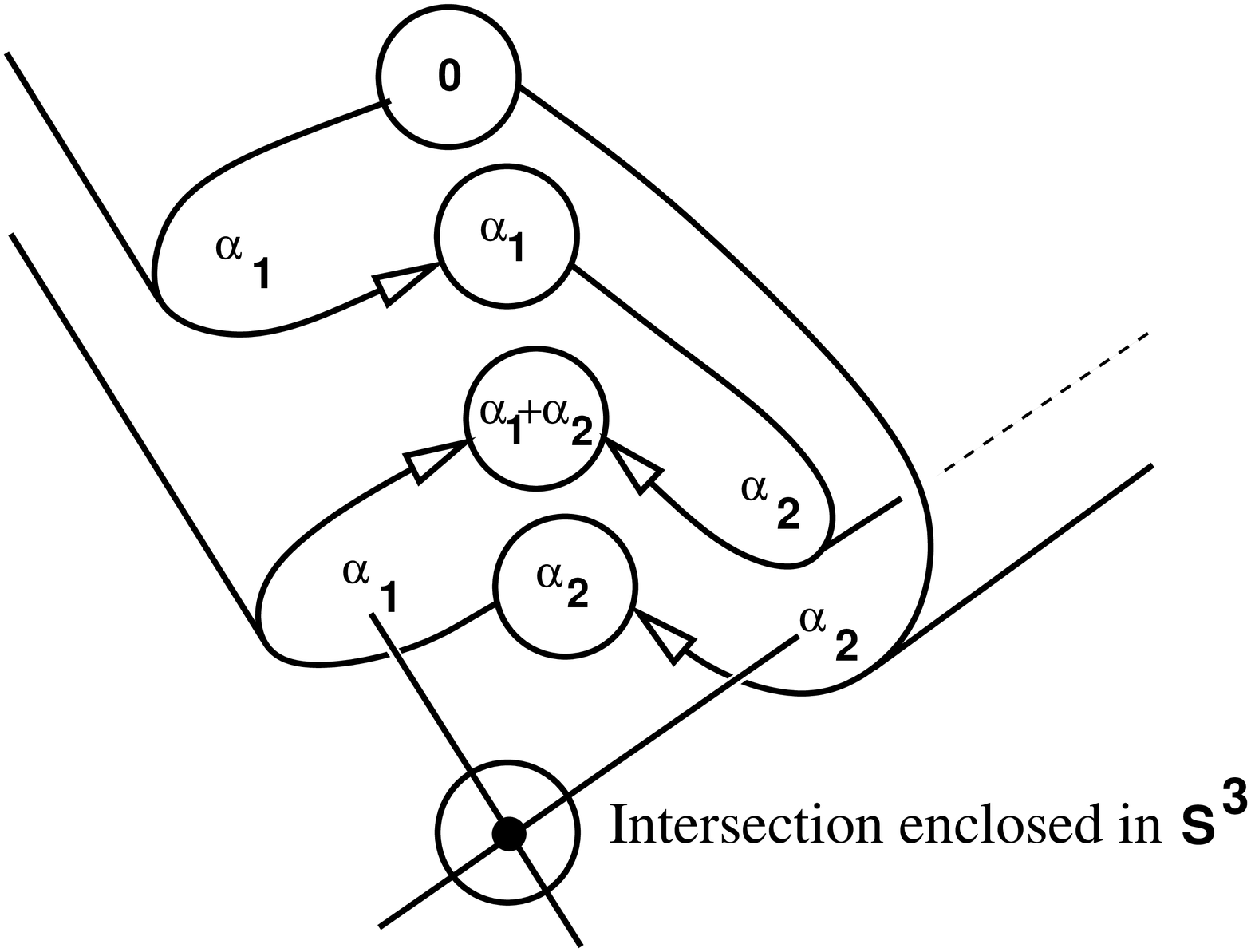}
\psfig{width=6cm, figure=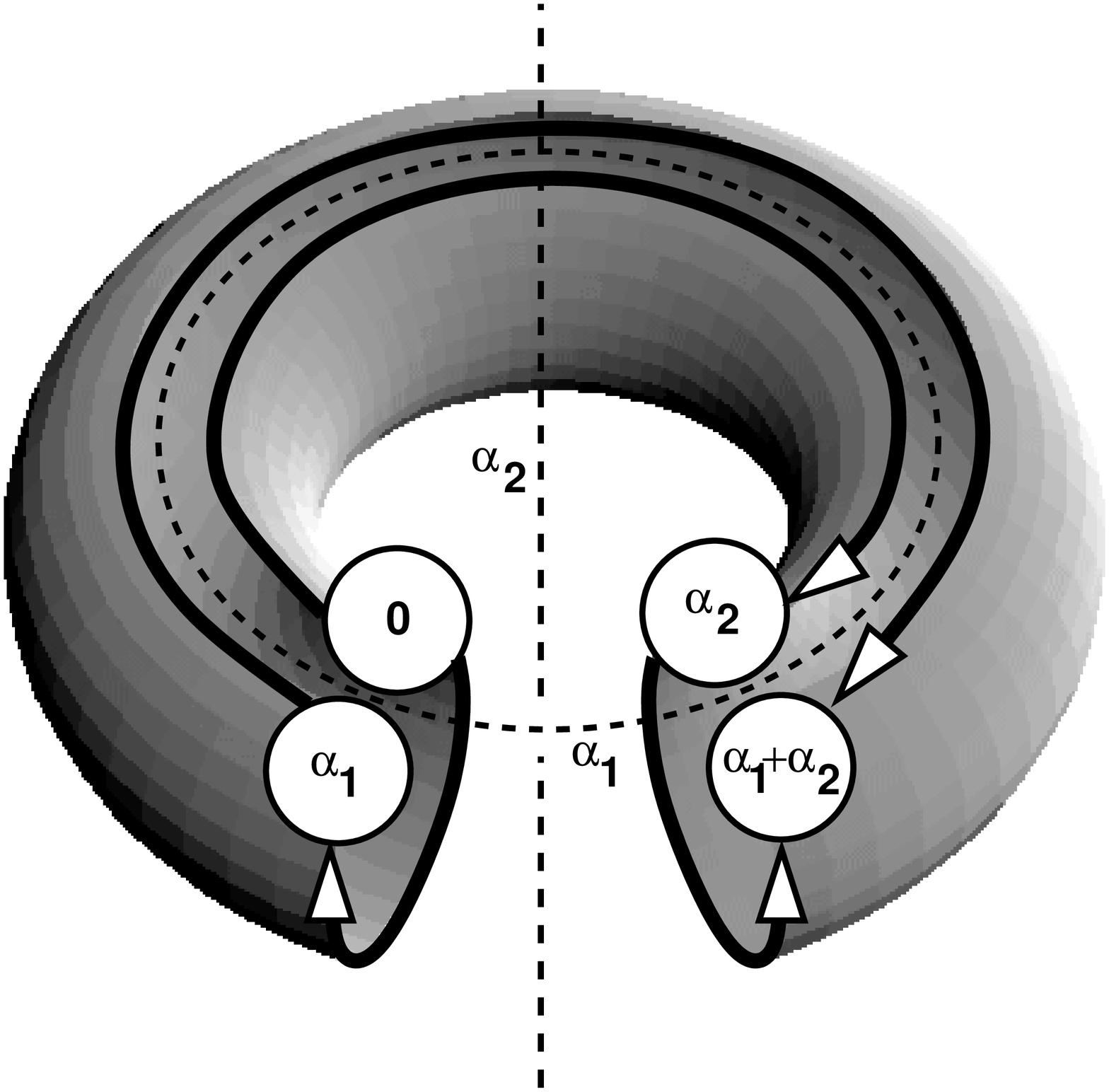}\\
\hspace{1.5in} (a) \hfill (b) \hspace{1.5in}
\caption [] {(a) The hole at a single corner.
(b) A hole-filling disk in $\Sphere{3}$ enclosing the intersection
point of the two (complex) branch lines.}
\label {b5commutator.fig}
\end {figure}

\item[Step 3.]  To show that one can fill in these holes, we observe
 that a small 3-sphere around an intersection of two branch lines,
 with the branch lines taken off, is homotopic to a torus and hence
 its fundamental group is isomorphic to the Abelian group $\mathbb Z
 \times \mathbb Z$. It is easy to see that the boundary of a small
 hole at this intersection, which lies in the surrounding 3-sphere,
 represents the element $(1,0)+(0,1)+(-1,0)+(0,-1)=(0,0)$ and
 therefore is contractible (a filled-in disk as illustrated in Figure
 \ref{b5commutator.fig}(b)).

{\it Remark\/}: If one tries to construct such a surface directly in
$\mathComplex^{2}$, then there would be a hole at a corner where
three branch lines intersect, and for such holes the argument above fails.

\end{itemize}

We have therefore obtained an oriented closed surface $F$ immersed in
$S$, whose Euler characteristic can be easily seen to be
$\chi(F)=40-80+32=-8$.  Hence the surface $F$ is of genus 5, i.e., a
{\it sphere with five handles\/}.

Now, let
\[
\epsilon(\alpha_1,\alpha_2,\alpha_3,\alpha_4,\alpha_5)=\int_F\,\beta_5(z_1,z_2;\,
  \alpha_1,\alpha_2,\alpha_3,\alpha_4,\alpha_5)\,dz_1\wedge dz_2 \ .
\]
Clearly $\epsilon(\alpha_1,\alpha_2,\alpha_3,\alpha_4,\alpha_5)$ is a holomorphic function of
$(\alpha_1,\alpha_2,\alpha_3,\alpha_4,\alpha_5)$ and is invariant
under continuous deformations of $F$.  As in the case of the  Pochhammer contour
described in  Section \ref{sec:B4Poch}, we can  take the limit and
 calculate for suitably restricted $(\alpha_1,\alpha_2,\alpha_3,\alpha_4,\alpha_5)$ that
\begin{eqnarray*}
\lefteqn{\epsilon(\alpha_1,\alpha_2,\alpha_3,\alpha_4,\alpha_5)=}\\
 && = (1-e^{2\pi i\alpha_1})(1-e^{2\pi i\alpha_2})(1-e^{2\pi
i\alpha_3})(1-e^{2\pi i\alpha_4})(1-e^{2\pi i\alpha_5}) \\
&& \!\!  \cdot  \!\! \iint\limits_{0<x_{2}<x_{1}<1}\,
 \!\!\! x_1^{\alpha_1-\alpha_2-\alpha_5}x_2^{\alpha_2-1}(1-x_1)^{\alpha_3-1}
  (x_1-x_2)^{\alpha_5-1}(1-x_2)^{\alpha_4-\alpha_3-\alpha_5}\,dx_1\,dx_2\, \ .
\end{eqnarray*}
This proves (\ref{B5ratio.eq}). $\Box$

Following the analogy to the $B_{4}$ Pochhammer analysis to determine
further constraints on the poles and zeroes of $B_{5}$ is an
interesting challenge for future work.

 \section{Visualizations of Connected Components and  Pochhammer Contours}
 \label{sec:Vis}

  The analysis of the $B_{5}$ function in the previous sections has
been based entirely on algebraic manipulations and line drawings
sketching the essential features of the geometry.  This section is
motivated by the observation that, since there are  algebraic
constructions for  every geometric concept, we can go one step further
and show precise images of each construction, helping the reader to
develop a quantitative as well as a qualitative understanding of the
framework we have developed.  We establish the basic context with
some examples based on the Euler Beta function, and then proceed to
show some of the remarkable manifolds that occur in the $B_{5}$ analysis.

\subsection{${\mathbf B_4}$ Connected Components Embedded in a
  Veronese Surface}

The Euler Beta function itself can be analyzed using cross-ratio
coordinates.  We begin with the two cross-ratio  variables, $x_1$ and $x_2$, obeying
the apparently uninteresting constraint
\[x_{1} = 1 - x_{2} \ . \]
However, when we put this into homogeneous coordinates
$\{x_{0},x_{1},x_{2}\}$, the constraint becomes $x_{0} = x_{1}
+x_{2}$, and we can solve these equations independently in the three component
regions, written as three intervals in inhomogeneous coordinates as  $A = [0,1]$, $B =
[1,\infty]$, and $C = [-\infty,0]$. Noting that the
space we are now dealing with is {\it not\/} $\,\mathComplex$ or $\CP{1}$,
but the real part of the $\CP{2}$ {\it cross-ratio-space\/},
we can parameterize each interval in homogeneous $\RP{2}$
coordinates as follows:
\begin{eqnarray}
A(t):&& \left[ 1,\, t,\, (1-t) \right] \nonumber\\
B(t):&& \left[ (1-t), \, 1,\, -t  \right] \\
C(t):&& \left[-t,\, (1-t),\, -1 \right] \nonumber \ .
 \end{eqnarray}
We see that region $A$ solves $1= x_1+x_2$ with $x_{1} = t$,
$B$ solves $1= x_1+x_2$ with $x_{1} = 1/(1-t)$ when all is multiplied
by $(1-t)$, and
$C$ solves $1= x_1+x_2$ with $x_{1} = (t-1)/t$ when all is multiplied
by $t$.  The interpolating functions in the three regions
obviously correspond to the three $B_4$ component regions introduced
initially in Figure (\ref{b4-3conn.fig}),
and they interpolate between the  points represented
in the Riemann-sphere depiction of Figure \ref{rsphere1.fig}:

\begin{figure}[!htb]
\centering \mbox{\psfig{width=8.5cm, figure=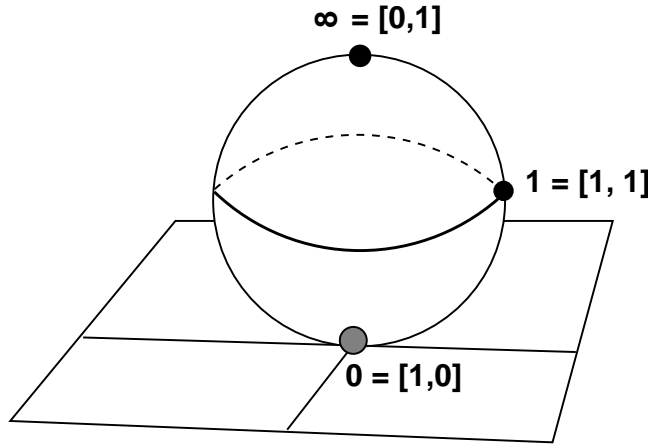}}
  \caption[]{The full base space of the $B_4$ branched covering, showing
the three regions corresponding to the three intervals $(0,1)$,
$(1,\infty)$, and $(-\infty,0)$ on the real projective  line.}
  \label{rsphere1.fig}
\end{figure}

\[
\begin{array}{rcrclclcl}
&&&& \RP{2} \mathrm{\ homog} &&
             \CP{1}  (z_{0},z_{1}) &&
            \mathComplex  \mathrm{\ inhomog\ } (z_{1}/z_{0})\\
p_{1} & = & A(0) & = & [1,0,1] & \approx &  z = [1,0]   & \approx &   x=0 \\
p_{2} & = & B(0) & = & [1,1,0] & \approx  & z = [1,1]  & \approx &   x=1 \\
p_{3} & = & C(0) & = & [0,1,-1] &\approx  & z = [0,1]& \approx &  x=\pm \infty \ .
\end{array}
\]

\begin{figure}[!phtb]  
\centerline{\psfig{width=6.5cm, figure=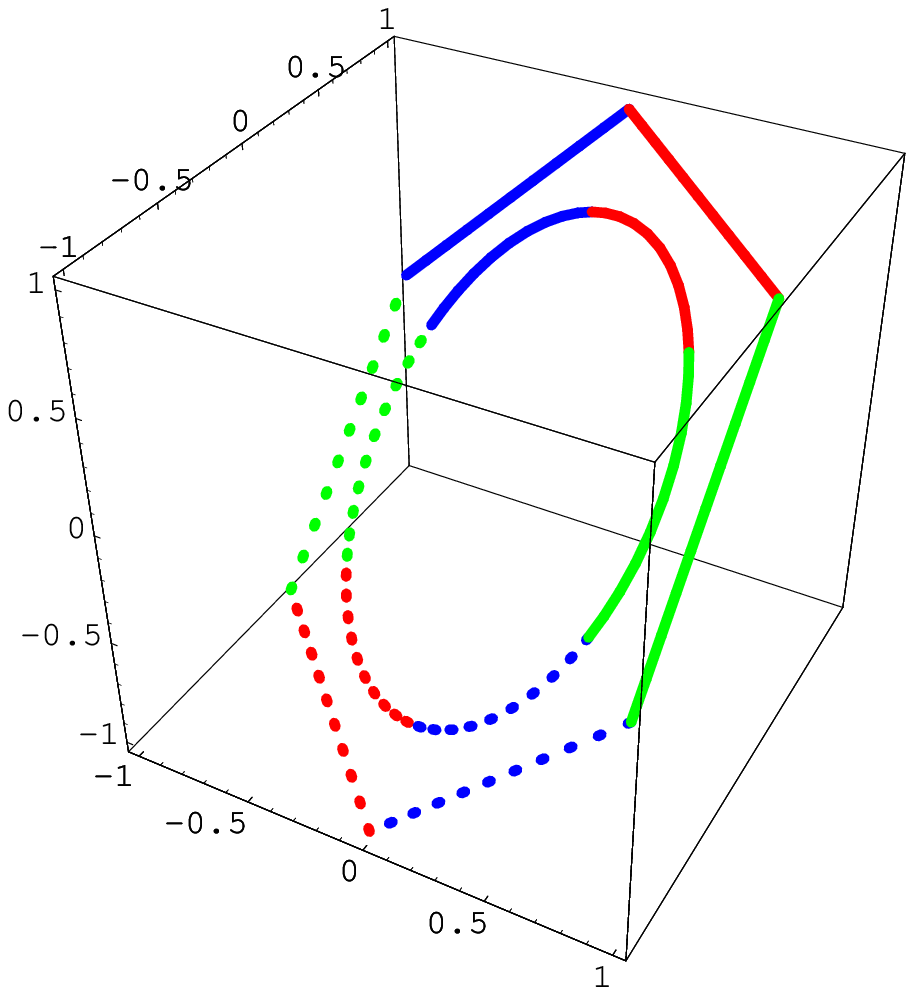}}
\centerline{ (a) }
\centerline{   \psfig{width=6.5cm, figure=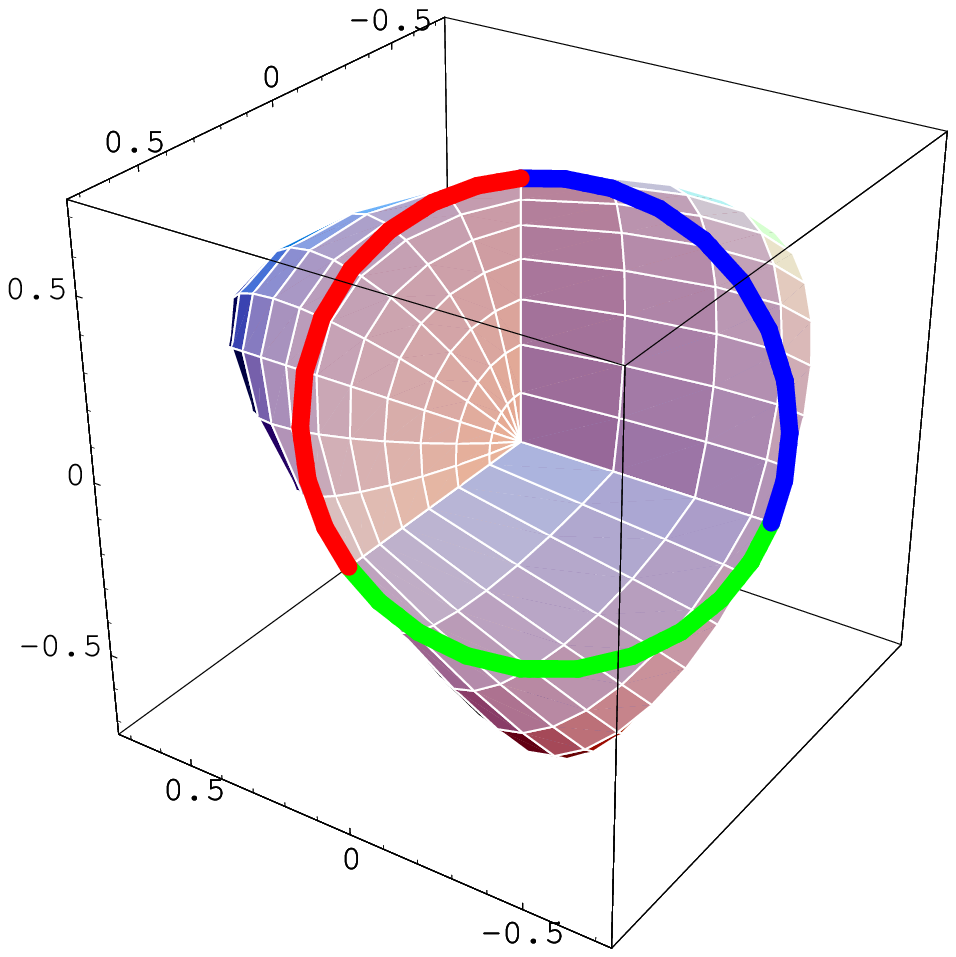}
\psfig{width=7.0cm, figure=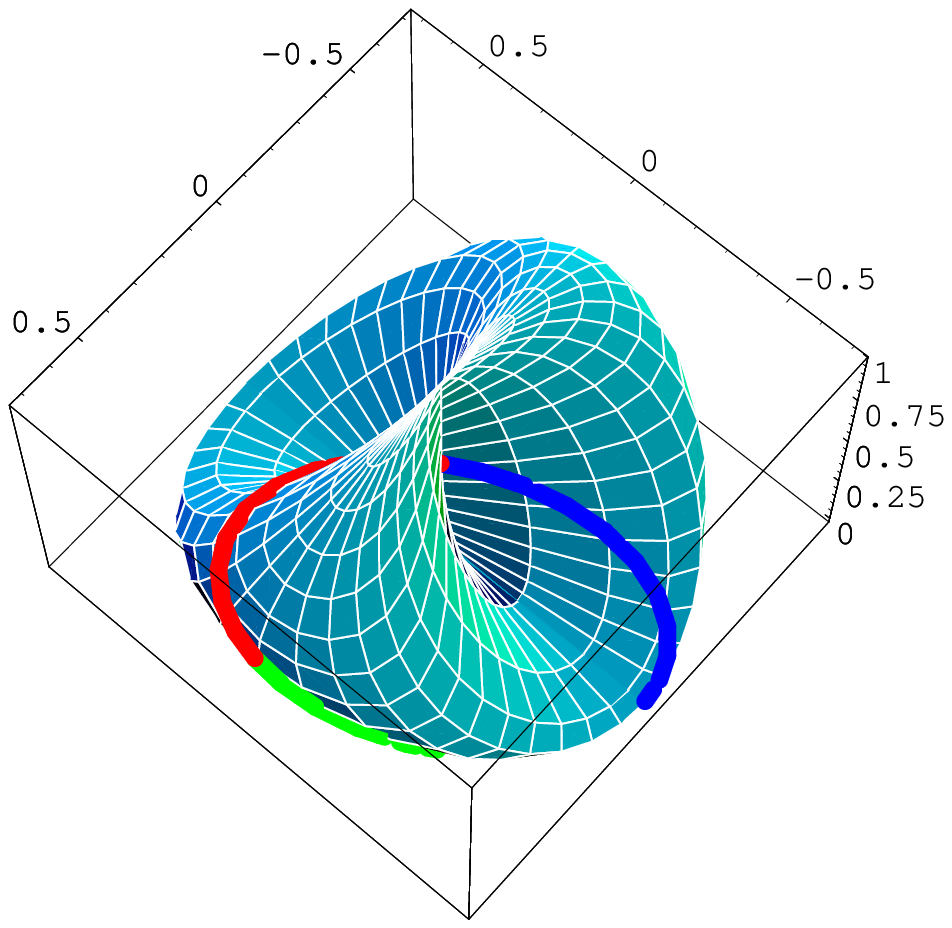}}
\centerline{\hspace{1.in}  (b)\hfill  (c)\hspace{1.in}}
  \caption[]{(a) The double cover of $B_4$ branched covering, showing
the linear interpolations among the points
$p_{1} = (1,0,1), \bar{p}_{1} = (-1,0,-1),
p_{2}  =  (1,1,0), \bar{p}_{2}  =  (-1,-1,0),
p_{3}  =  (0,1,-1), \bar{p}_{3}  =  (0,-1,1)$
 and their projections to $\Sphere{2}$.  (b) The Veronese map, projected
 to the Steiner Roman Surface, $(w_1,w_2,w_3)$. (c) Projection onto the
crosscap, $(w_1,w_2,w_6)$.}
  \label{b4-double-rp2.fig}
\end{figure}

But there is a small problem: if we follow the coordinate
interpolations carefully, they only work projectively; the actual
interpolations close on one another only if we include the negative,
projectively-equivalent points $\bar{p}_{i} = - p_{i}$, for a total
of 6 points and six linear paths, rather than three.  Thus, the
projective coordinates for the three components  $A,\,B,\,C$
can be plotted either as a connected hexagon in $\mathReal^{3}$, the
embedding space of the homogeneous $\RP{2}$ coordinates, or as a more
visually consistent projection onto a constant-radius $\Sphere{2}$,
the double-cover of $\RP{2}$, as shown in Figure \ref{b4-double-rp2.fig}(a).

To actually achieve the desired end result of a visualization of
the $B_{4}$ cross-ratio coordinates in a logical embedding, we must
find a quadratic map that removes the distinction between
the positive and negative versions of the same projective coordinates
and maps $\Sphere{2}$ explicitly to $\RP{2}$.  This is achieved
classically by the Veronese surface (see, e.g., the traditional
embedding of $\RP{2}$ given in the appendix
of Hilbert and Cohn-Vossen, \cite{HilbertCV}):
\begin{equation}
\begin{array}{ccc}
w_1  = \sqrt{2} x_{0} x_{1} &
w_2  = \sqrt{2} x_{0} x_{2} &
w_3  = \sqrt{2} x_{1} x_{2} \\
w_4  = ( x_{0})^{2} &
w_5  = ( x_{1})^{2} &
w_6  = ( x_{2})^{2}
\end{array}
\label{b4vero.eq}
\end{equation}
where the spherical constraint $x_{0}^{\ 2} + x_{1}^{\ 2} + x_{2}^{\ 2} =1$
implies the standard Veronese surface constraint  $\sum (w_{i})^{2} =
\left(\sum(x_{k})^{2}\right)^{2} = 1$.
In  Figures \ref{b4-double-rp2.fig}(b) and (c), we see the exact
paths of the three component integrals of the Euler Beta function
as they are embedded in alternate projections of $\RP{2}$ (the
real part of $\CP{2}$) to 3D.
This is equivalent mathematically, and yet a significantly
contrasting viewpoint, to the conventional $\CP{1}$ alternative indicated in
Figure \ref{rsphere1.fig}.

\subsection{Visualizing the ${\mathbf B_4}$ Pochhammer Contour}

\begin{figure}[!htbp]
{\centering \vspace*{-1in}
  \mbox{\psfig{width=7.0cm, figure=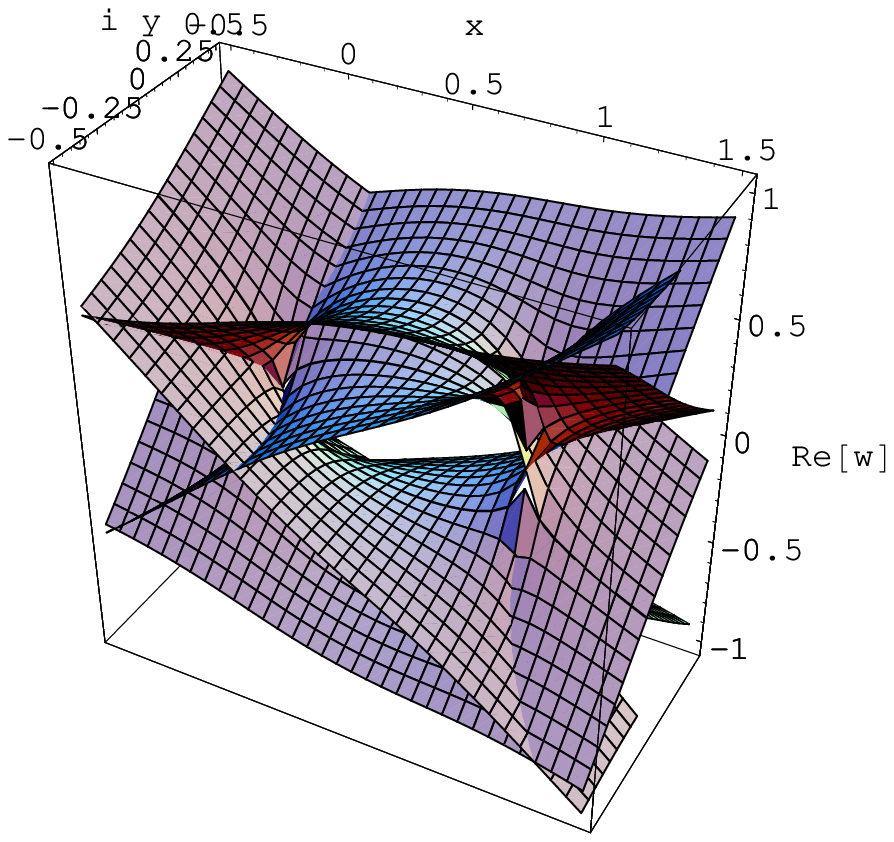}
\psfig{width=7.0cm, figure=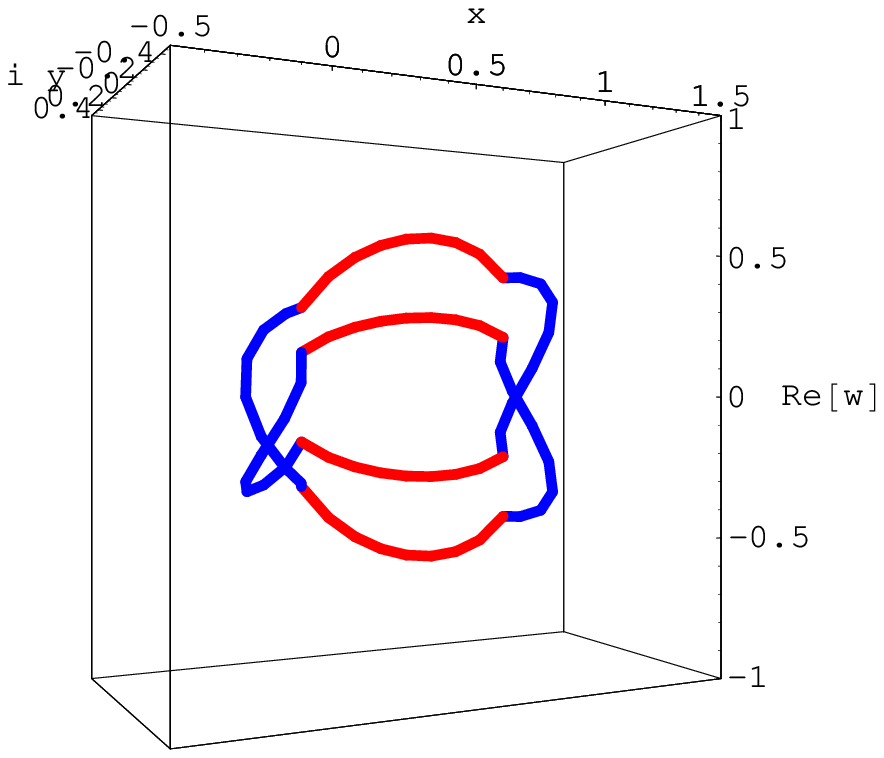}}\\
 \mbox{(a) \hspace{3.0in} (b)}  \\  
\mbox{\psfig{width=9.0cm, figure=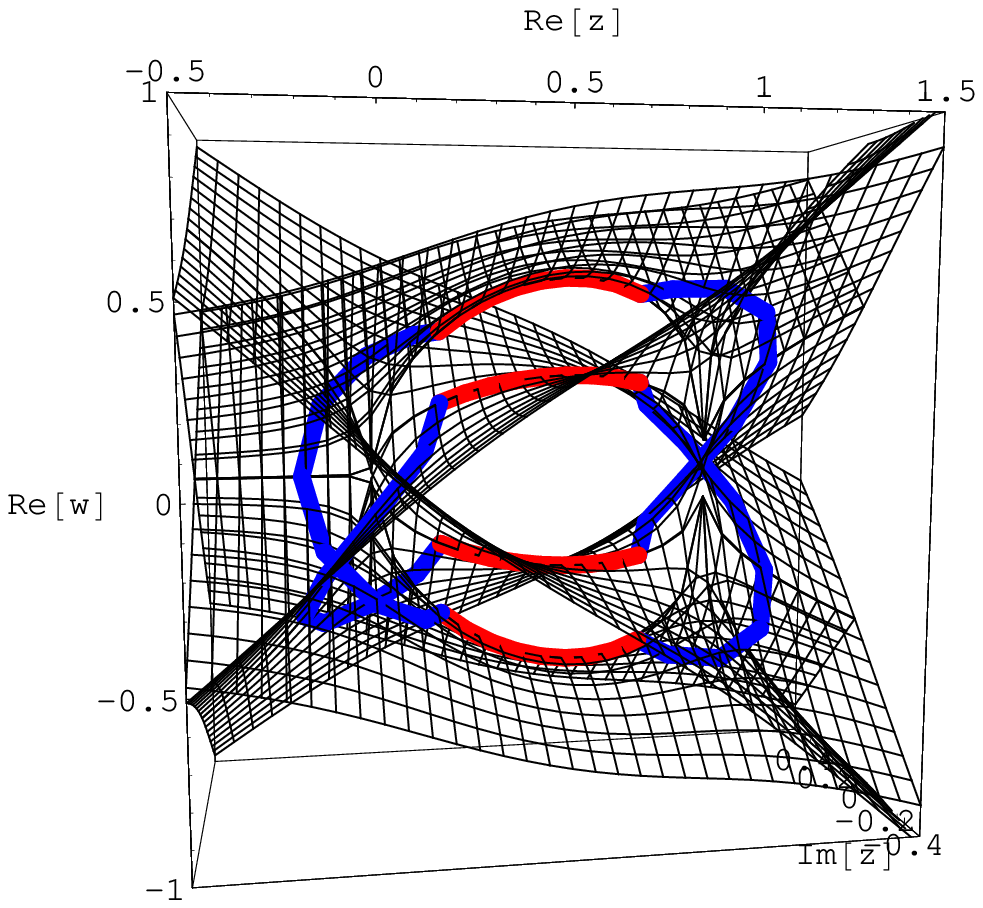}}\\
\hfill (c) \hfill }
  \caption[]{(a) A representative Riemann surface derived from the
integrand of the Euler Beta function with suitable rational values of
$(\alpha_{1},\,\alpha_{2})$.  (b) The actual geometry of the Pochhammer contour traced on
the representative Riemann surface.  The path on the Riemann surface
is a ``commutator,'' encircling each branch point once in each
direction; comparing to Figure \ref{pochsketch.fig}, the four end
loops shrink to points as $r \rightarrow 0^{+}$. (c) Combined plot. }
  \label{pochviews.fig}
\end{figure}

  We now illustrate explicitly the geometry of the Pochhammer contour for the
Euler Beta function.  Starting from (\ref{b4integrand.eq}), we choose
a pair of small relatively prime rational exponents $(\alpha_1,\,\alpha_2)$, and project
the 4D plot of $w=\beta(z;\alpha_1,\alpha_2)$ to 3D, with the horizontal plane
parameterized by $x={\rm Re}(z)$, $y={\rm Im}(z)$, and the vertical axis given
by ${\rm Re}(w)$.  Figure \ref{pochviews.fig}(a) shows a small region
of the branched Riemann cover of the complex plane punctured at $z=0$
and $z=1$,  and   Figure \ref{pochviews.fig}(b) shows the
precise path in this branched cover of the Pochhammer ``commutator''
contour sketched in Figure \ref{pochsketch.fig}, but now as an actual embedding in
$\mathComplex^{2}$ (technically $\mathReal^{4}$ projected to $\mathReal^{3}$).
Figure \ref{pochviews.fig}(c) combines the two views to show the
Pochhammer contour in its geometric context on the Riemann surface.

\subsection{${\mathbf B_5}$: Components of the 5-Point Cross-Ratios}

The four single lines joined by infinitesimal loops
shown in Figures \ref{pochsketch.fig} and
\ref{pochviews.fig} represent the four distinct {\it phases\/}
of the $B_{4}$ Pochhammer integration path.
For $B_{5}$, the analog of one of these lines is
a pentagonal surface, and the set of four lines representing integration domain
of $B_{4}$ with distinct phases is replaced in $B_{5}$ by 32
pentagons with distinct phases.  Just as the three lines in
Figure \ref{rsphere1.fig} or Figure
 \ref{b4-double-rp2.fig} describe the three {\it components\/}
of $B_{4}$ that followed from solving the cross-ratio constraint, the
twelve components of $B_{5}$ can be studied using parametric solutions
of its own cross-ratio constraint system:  re-indexing for convenience using
 $(z_{1},z_{2},z_{3},z_{4},x_{5})\equiv
(u_{12},u_{13},u_{23},u_{24},u_{34})$,
the $B_{5}$ cross-ratio system becomes:
\begin{eqnarray}
z_{1} &=& 1 - z_{3}z_{4} \nonumber \\
z_{2} &=& 1 - z_{4}z_{5} \nonumber\\
z_{3} &=& 1 - z_{5}z_{1} \label{b5constrnts.eq} \\
z_{4} &=& 1 - z_{1}z_{2} \nonumber\\
z_{5} &=& 1 - z_{2}z_{3}  \nonumber \ .
\end{eqnarray}

\begin{figure}[!ht]
\centering \mbox{\psfig{width=2.5in, figure=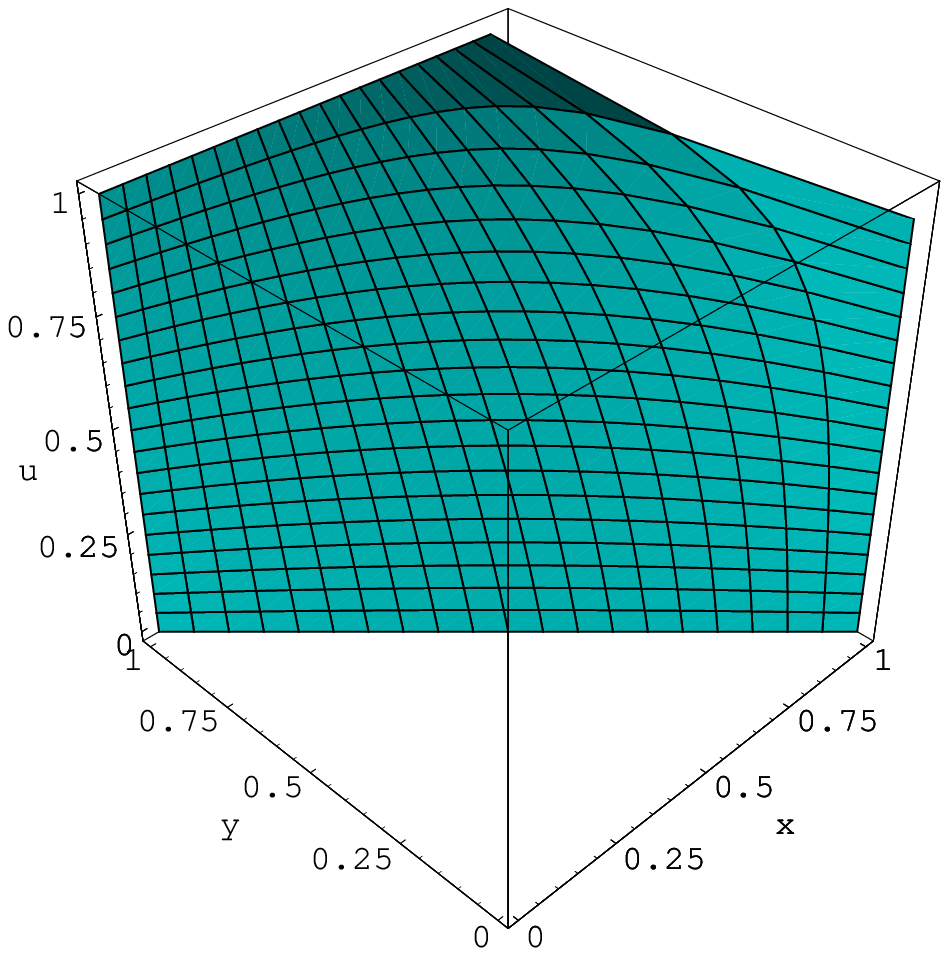}}
\mbox{\psfig{width=2.5in, figure=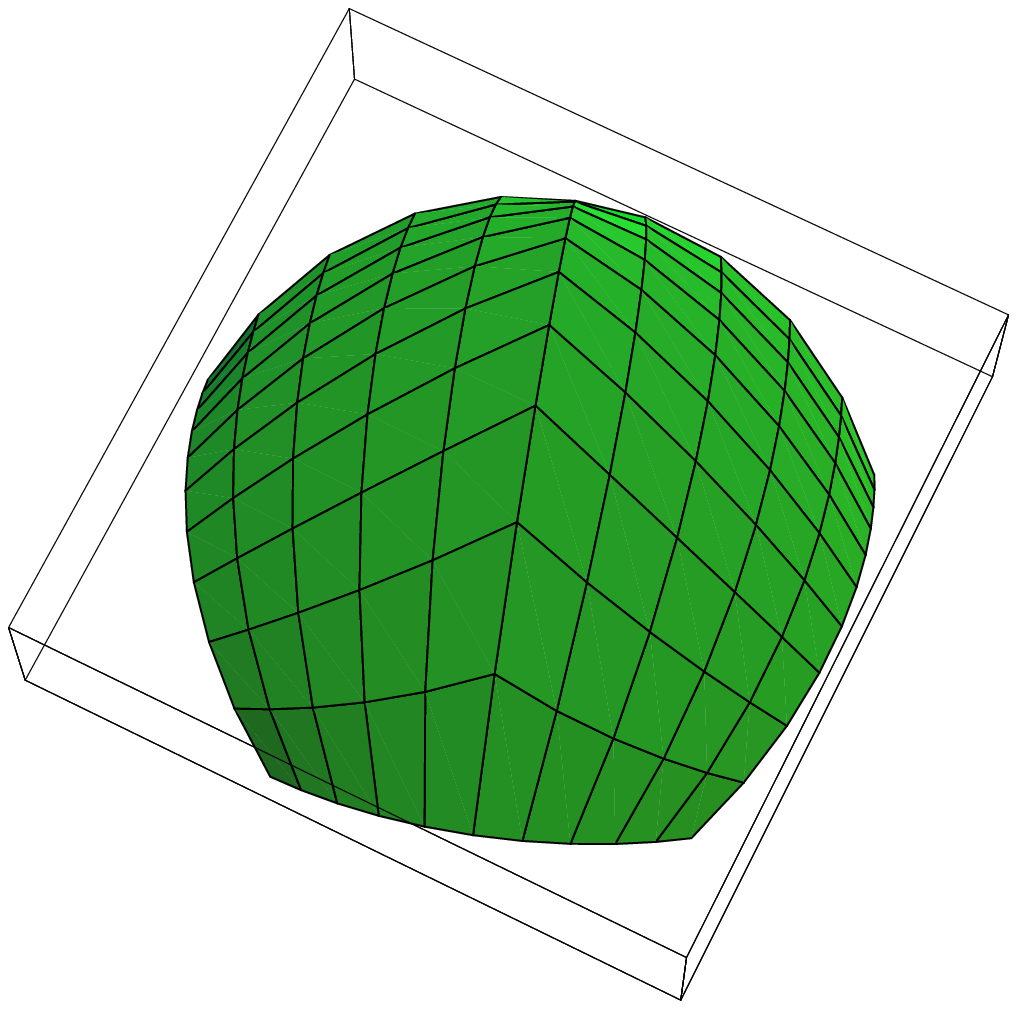}}\\
\centerline{\hspace{1.5in}(a) \hfill (b) \hspace{1.5in}}
\centerline{\psfig{width=3.5in, figure=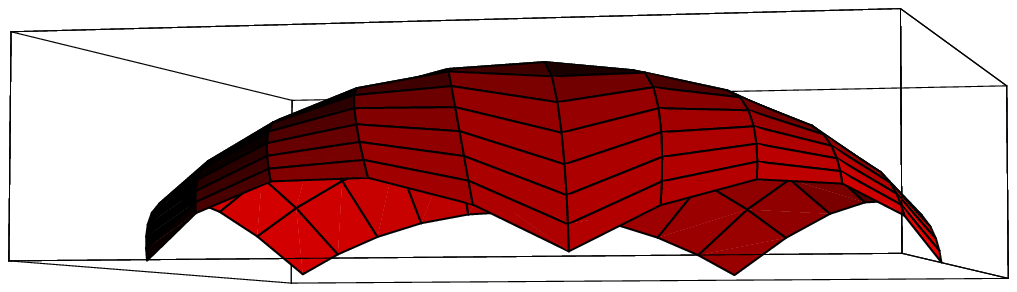}}
\vspace{0.1in}
\centerline{ (c) }
  \caption[]{(a) Plotting the $B_5$ variables $(x,(1-x)/(1-xy),y)$,
showing how the ``blown-up'' pentagonal manifold arises naturally
in the cross-ratio manifold. (b) Completely regular version of the
pentagonal, normalized to $\Sphere{5}$ and projected,
corresponding to the variables of  Figure \ref{uvmap.fig}.
(c) Side view of (b).}
  \label{pentplotmesh.fig}
  \end{figure}

  Any pentagon can be represented algebraically by picking
two of the five $z_{i}$'s as independent, and plotting any of the
dependent variables found by solving the constraints in formula
(\ref{b5constrnts.eq}) on the third axis.  The typical result, shown in Figure
\ref{pentplotmesh.fig}, is an algebraic 2-manifold embedded in
$\mathReal^{5}$ showing the integration
region over the variety given by (\ref{b5constrnts.eq}).
 Projected from a horizontal direction, the pentagon of Figure
\ref{pentplotmesh.fig}(a) becomes a square region, whereas when
projected from the vertical direction, it becomes a triangular region,
corresponding to formula (\ref{triB5.eq}).

\begin {figure}[tp!]
\centering
  \mbox{\psfig{width=4.5in, figure=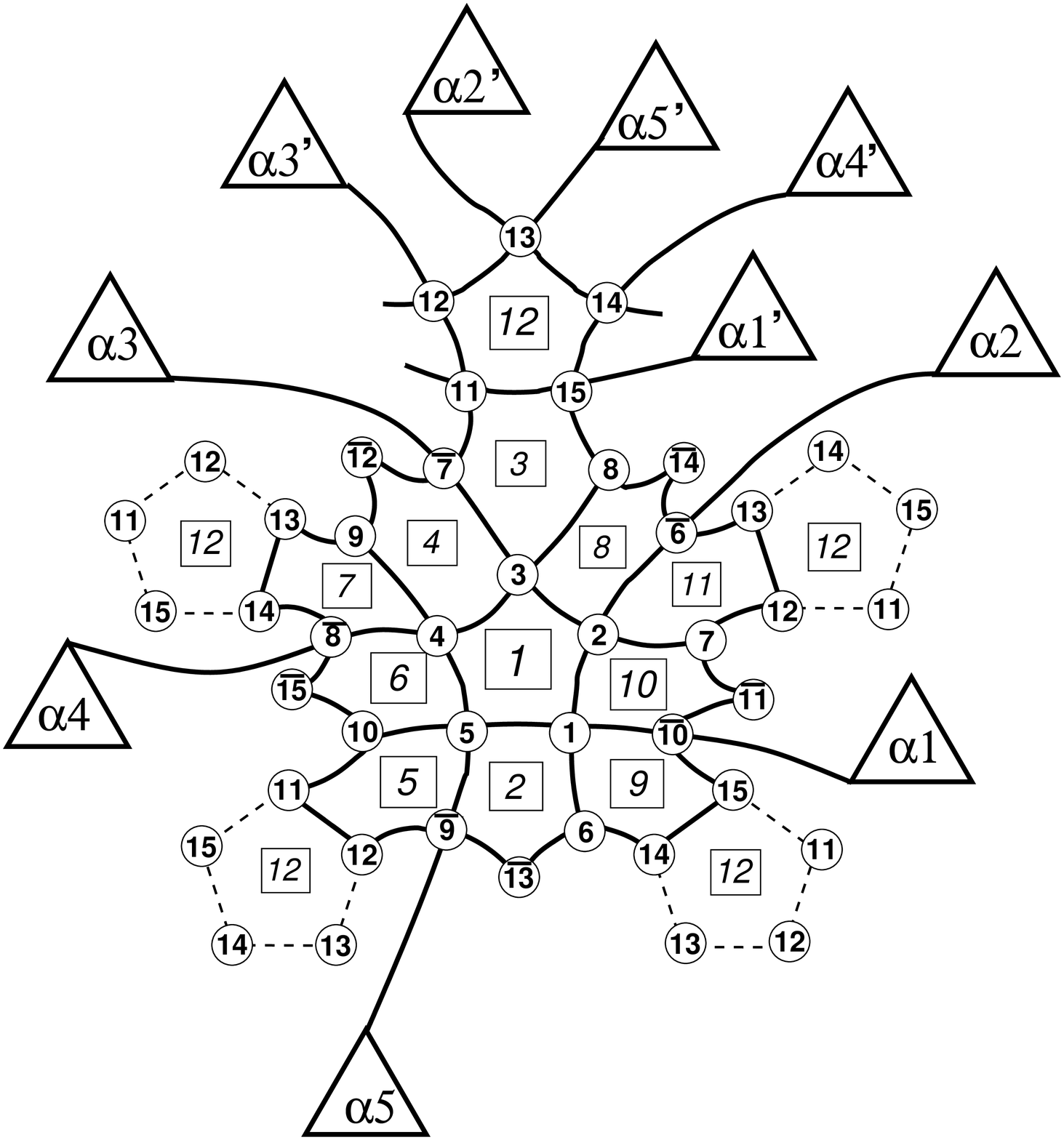}}
\caption [] {The diagram of how the 12 pentagonal pieces of surface
  join together to form a closed surface in $\RP{5}$.  Circles mark
  the 15 vertices, and squares mark the images of the corresponding
  regions in Figure \ref{12st.fig}.  Triangles label the exponent
of each of the branch lines delineating the connected components.
(See (\ref{b5expon.eq}).)}
\label {b5surfconnect.fig}
\end {figure}

\begin {figure}[tb]
\parbox[t]{5.00in}{\centering
  \psfig{file=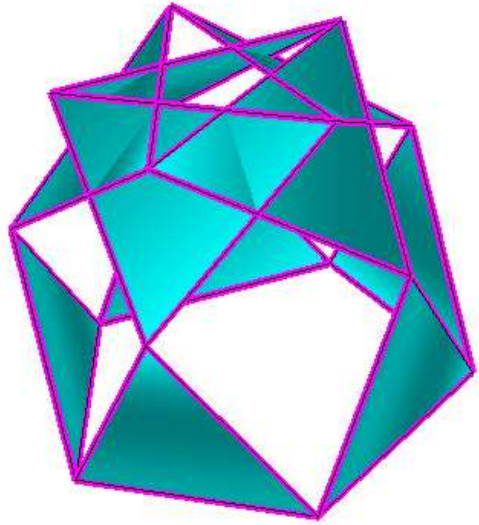,height=4.00in}
\caption[]{These 10 filled triangles represent the single
cover of the 10 $\CP{1}$
branch surfaces of the $B_{5}$ integrand Riemann manifold,
 The straight edges taken 5 at a time bound the 12 pentagons.}
\label{pentTris.fig}}
\end{figure}

To create an image of the twelve $B_{5}$ components, we now use the
constraints (\ref{b5constrnts.eq}) and proceed through the same
arguments that we used for $B_{4}$: We solve the constraints in
a family of homogeneous $\RP{5}$ coordinates based on choosing
non-singular parameterizations of
$(z_{0},z_{1},z_{2},z_{3},z_{4},z_{5})$, and find that the natural
connectivity actually gives us initially the 24-pentagon double cover
analogous to the six $B_{4}$ curves shown in Figure
\ref{b4-double-rp2.fig}(a).  Figure \ref{b5surfconnect.fig} is the
schematic analog of Figure \ref{rsphere1.fig} for $B_{4}$, showing
 the topological diagram of the
surface, which we can verify is non-orientable with 15 vertices, 30
edges, and 12 pentagonal faces, giving the advertised Euler
characteristic $\chi = -3$,  a sphere with five crosscaps.  However,
we can also see traced on this surface the family of complex lines
that form the symmetrized base of the branched cover enabled by the
blow-ups: there are 10 separate interlocking triangles, each denoting
the (circular) real
line of a $\CP{1}$ corresponding precisely to the $B_4$ diagram of
Figure \ref{rsphere1.fig} or  \ref{b4-double-rp2.fig}(b);
treating these schematically as  filled-in
triangles, we get the image in Figure \ref{pentTris.fig}, where the
boundaries of the 10 triangles taken five at a time bound the 12
pentagons.  The corresponding 12-pentagon figure can be thought
of as shown in Figure \ref{b5surfconnect.fig}, where the boundaries of the connected
components (corresponding to $-\alpha_{1}-\alpha_{2}$ for $B_{4}$)
are linear combinations of the $B_{5}$ $\alpha_{i}$'s.
In particular, the face labeled ``12'' corresponds to a $B_5$ function with
a set of exponents that is distinct from the values
$(\alpha_1,\alpha_2,\alpha_3,\alpha_4,\alpha_5)$  used in
(\ref{triB5.eq}), although they are closely related.
One can show with suitable variable changes that the exponents
corresponding to the primed branch lines (analogous to the exponent at
infinity for $B_{4}$) are
\begin {eqnarray}
{\alpha'}_1 & = & 1 + \alpha_{1} - \alpha_{2} - \alpha_{5} \nonumber \\
{\alpha'}_2 & = & 1 + \alpha_{2} - \alpha_{3} - \alpha_{1}  \nonumber \\
{\alpha'}_3 & = & 1 + \alpha_{3} - \alpha_{4} - \alpha_{2} \label {b5expon.eq}\\
{\alpha'}_4 & = & 1 + \alpha_{4} - \alpha_{5} - \alpha_{3}  \nonumber \\
{\alpha'}_5 & = & 1 + \alpha_{5} - \alpha_{1} - \alpha_{4}  \nonumber \ .
\end{eqnarray}

\begin {figure}[tbp]
\parbox[t]{5.00in}{\centering
  \psfig{file=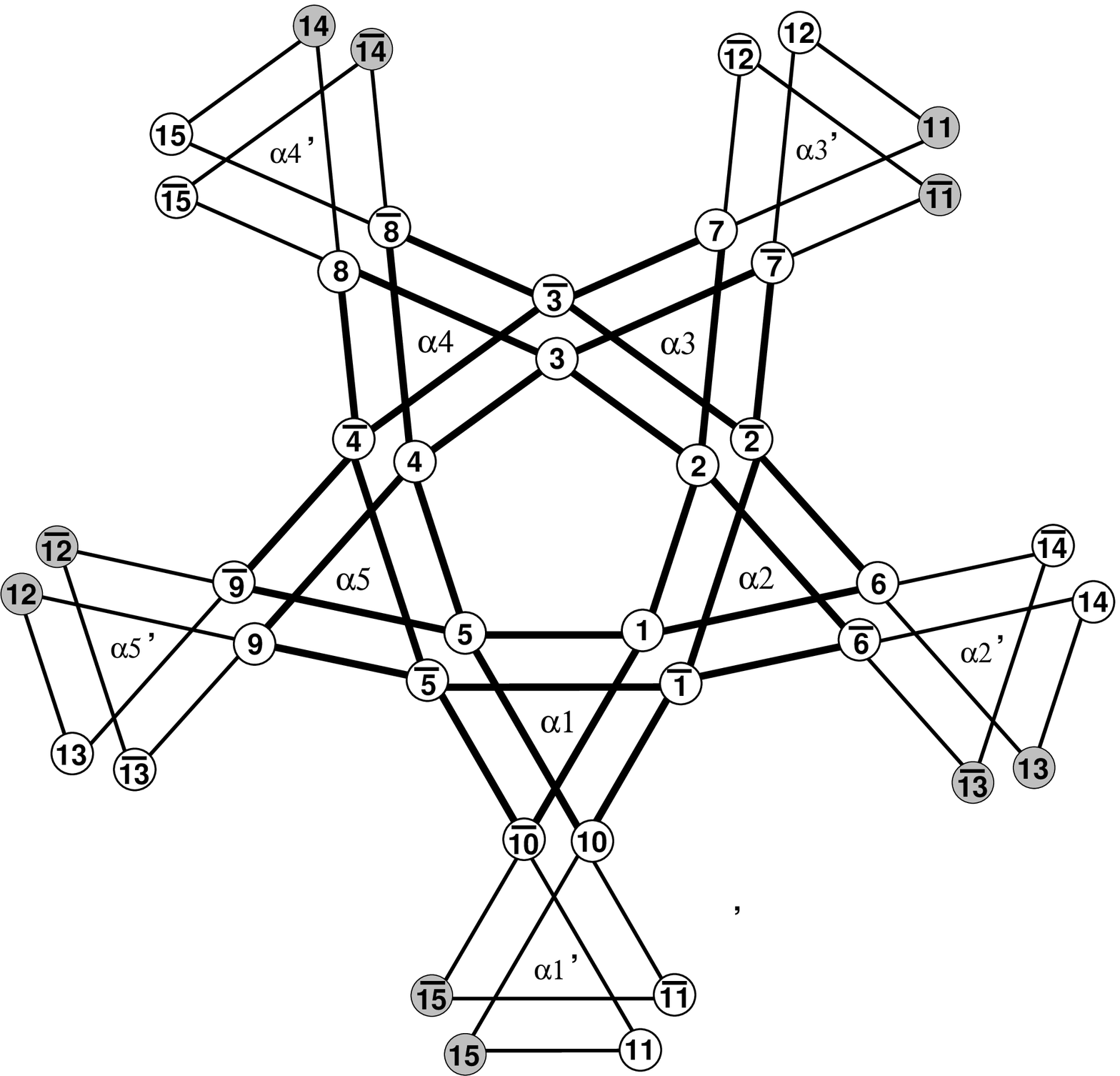,height=4.00in}
\caption[]{An explicit graph of the double covering of the
cross-ratio variable space, showing each vertex
and each of the 10 hexagons corresponding individually
to a double-covered $\CP{1}$ branch line, and also to
the hexagon in Figure \ref{b4-double-rp2.fig}. It is easy to
see how Figure \ref{pentTris.fig} emerges as the single-cover after
reducing each hexagon to a triangle.}
\label{pentHex.fig}}
\end{figure}

The solutions of the constraints are continuous only in the double
cover, so we will work first in an unnormalized $\RP{5}$ to produce
the analogs of the six end points and six straight interpolating edges that we
showed in Figure  \ref{b4-double-rp2.fig}(a) for $B_{4}$.
This equivalent set of vertices is the set of 10 hexagons representing
the double cover of the $\CP{1}$ branch lines denoted by
$(\alpha_{1},\ldots,{\alpha'}_{1},\ldots)$, as shown in Figure \ref{pentHex.fig}.
These  have the following
vertex assignments in the double cover:
\[
\begin{array}{rlrl}
\mbox{\rm line\ } \alpha_{1}: & (1, 5, 10,-1,-5,-10) &
\mbox{\rm line\ }  {\alpha'}_{1}: &(10,15,11,-10,-15,-11)\\
\mbox{\rm line\ } \alpha_{2}: & (2,1,6,-2,-1,-6)  &
\mbox{\rm line\ }  {\alpha'}_{2}: & (6,13,14,-6,-13,-14)\\
\mbox{\rm line\ } \alpha_{3}: &  (3,2,7,-3,-2,-7)&
\mbox{\rm line\ }  {\alpha'}_{3}: & (7,11,12,-7,-11,-12)\\
\mbox{\rm line\ } \alpha_{4}: & (4,3,8,-4,-3,-8)  &
\mbox{\rm line\ }  {\alpha'}_{4}: &  (8,14,15,-8,-14,-15)\\
\mbox{\rm line\ } \alpha_{5}: &  (5,4,9,-5,-4,-9)&
\mbox{\rm line\ }  {\alpha'}_{5}: &(9,12,13,-9,-12,-13) \ .
\end{array}
\]
Each region is bounded by a linear interpolation connecting the
(doubled) vertex set, as noted earlier.

\begin {figure}[tp]
\centering
\psfig{width=2.5in, figure=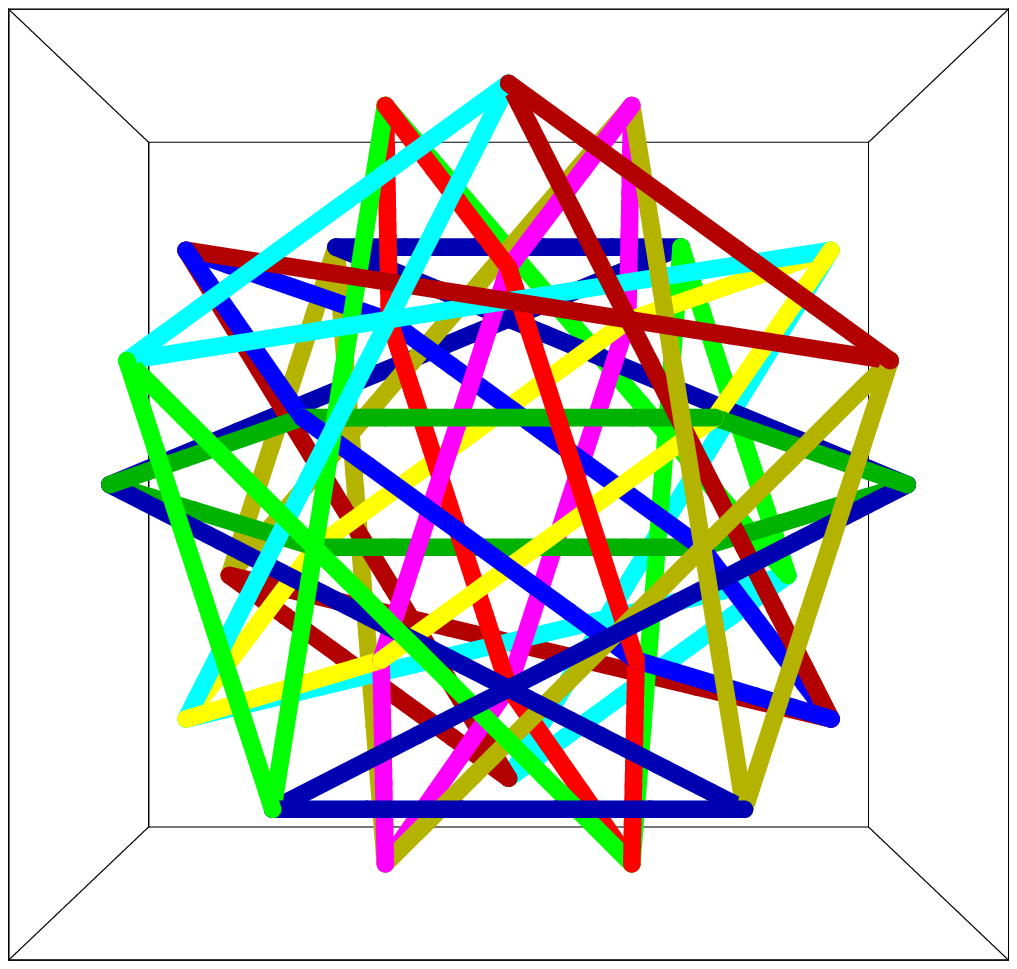}
\psfig{width=2.5in, figure=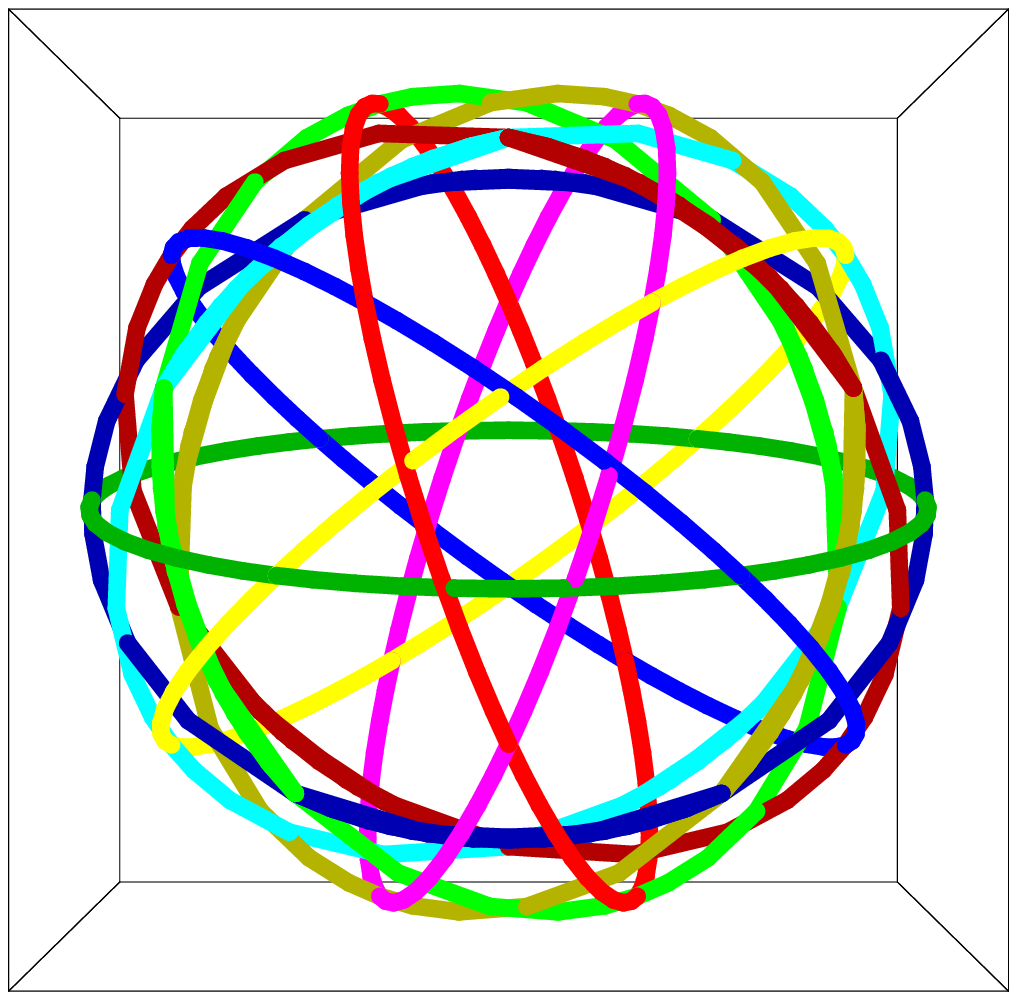}\\
\hspace{1.25in} (a) \hfill (b) \hspace{1.25in}
\caption [] {(a) Projection from homogeneous $\mathReal^{6}$
coordinates of all the  $2\times 30 = 60$
straight edges giving edges of pentagons (or
real lines of the $\CP{1}$ branches) in the double cover, the analogs
of the six {\it end points\/} in Figure  \ref{b4-double-rp2.fig}(a).
(b) Normalization to $\Sphere{5}$, analogous to the same six
points on $\Sphere{2}$ in  \ref{b4-double-rp2.fig}(a).}
\label{b5cp1doubles.fig}
\end{figure}

 Figure \ref{b5cp1doubles.fig} shows the actual doubled geometry,
both as straight lines in $\mathReal^{6}$, and as
curves in the sphere $\Sphere{5}$ (projected to 3D).

\paragraph{Fully Symmetric Vertex Choice.}
  If all we were interested in was the descriptive topology of the
$B_5$ five-crosscap base manifold, any set of vertices with the proper
connectivity would be sufficient.  However, our dual purpose is to
understand not only the topology, but also any unique geometric
features or symmetries that might characterize this manifold, leading
to embeddings whose graphical depictions might be especially
informative.

  We have therefore pursued the search for special
embeddings one step further, and computed an orthonormalized set
of vertices along with the corresponding surface embedding that allows all
pentagons to be expressed as rigid transformations of one another
derived from the operations of the discrete symmetry operators of
Section \ref{sec:symmetries}.

 The next step is to use the matrix $P$ defined by (\ref{PtPeqQ.eq}).
Such a $P$ can be found from standard linear algebra methods; we will
not write $P$ explicitly because its entries are not rational numbers
and are very lengthy.


Let
\begin{eqnarray*}
  F^{\pm}(s,t)& =& P \cdot \tilde{f}^{\pm}(s,t) \\
     & = & \pm \frac{P\cdot f_{i}}{\sqrt{f^{t}_{i} \cdot Q \cdot f_{i}}}
\end{eqnarray*}
We then get the transformation $P\cdot \widetilde{M}$ of the surface
$\widetilde{M}$.  Notice that  $P\cdot \widetilde{M}$ is in
$\Sphere{5}$ and is invariant under
\[  \gamma = P \cdot g \cdot P^{-1} \]
for $g\in\widetilde{G}$;  the $\gamma$'s are now the orthogonal matrices
forming a subgroup of order 240 in $\OO{6}$.

We pick the following 24 elements from  this group:
\begin{scriptsize}
\begin{equation}
\begin{array}{cc}
\gamma^{\pm}_1 = \pm {\mbox{Identity}} &
 \gamma^{\pm}_2 = \pm P\cdot  \left[ \begin{array}{cccccc}
 1&0&0&-1&0&0 \\
 2&0&-1&-1&0&-1\\
 0&0&0&-1&0&0\\
 2&-1&0&-1&-1&0\\
  0&0&0&0&0&1\\
 0&1&0&0&0&0
  \end{array} \right] \cdot P^{-1}\\[0.50in]
\gamma^{\pm}_3 = \pm P\cdot   \left[ \begin{array}{cccccc}
  -1&0&0&1&0&1 \\
  0&0&0&0&0&1 \\
  2&-1&-1&0&-1&0 \\
  0&0&0&1&0&0 \\
  -2&0&1&1&0&1 \\
  -2&1&0&1&0&1
  \end{array} \right] \cdot P^{-1} &
 \gamma^{\pm}_4 = \pm P\cdot   \left[\begin{array}{cccccc}
1& 0& 0& 0& 0& -1 \\
 2& -1& 0& -1& 0& -1 \\
 0& 0& 1& 0& 0& 0 \\
 0& 0& 0& 1& 0& 0 \\
 2& 0& -1& 0& -1& -1 \\
 0& 0& 0& 0& 0& -1
 \end{array} \right] \cdot P^{-1}\\[0.50in]
\gamma^{\pm}_5 = \pm P\cdot    \left[\begin{array}{cccccc}
-1& 1& 0& 1& 0& 0 \\
 -2& 1& 0& 1& 0& 1 \\
 -2& 1& 0& 1& 1& 0 \\
 0& 0& 0& 1& 0& 0 \\
 2& 0& -1& 0& -1& -1 \\
 0& 1& 0& 0& 0& 0
 \end{array} \right] \cdot P^{-1}&
 \gamma^{\pm}_6 = \pm P\cdot   \left[\begin{array}{cccccc}
1& -1& 0& 0& 0& 0 \\
 0& -1& 0& 0& 0& 0 \\
 2& -1& -1& 0& -1& 0 \\
 0& 0& 0& 1& 0& 0 \\
 0& 0& 0& 0& 1& 0 \\
 2& -1& 0& -1& 0& -1
 \end{array} \right] \cdot P^{-1}\\[0.50in]
\gamma^{\pm}_7 = \pm P\cdot    \left[\begin{array}{cccccc}
-1& 0& 1& 0& 0& 1 \\
 -2& 0& 1& 1& 0& 1 \\
 0& 0& 1& 0& 0& 0 \\
 2& -1& 0& -1& -1& 0 \\
 0& 0& 0& 0& 0& 1 \\
 -2& 0& 1& 0& 1& 1
\end{array} \right] \cdot P^{-1}&
 \gamma^{\pm}_8 = \pm P\cdot   \left[\begin{array}{cccccc}
1& 0& 0& 0& -1& 0 \\
 0& 1& 0& 0& 0& 0 \\
 0& 0& 1& 0& 0& 0 \\
 2& -1& 0& -1& -1& 0 \\
 0& 0& 0& 0& -1& 0 \\
 2& 0& -1& 0& -1& -1
 \end{array} \right] \cdot P^{-1}\\[0.50in]
\gamma^{\pm}_{9} = \pm P\cdot   \left[ \begin{array}{cccccc}
-1& 1& 0& 0& 1& 0 \\
 0& 0& 0& 0& 1& 0 \\
 -2& 1& 0& 1& 1& 0 \\
 -2& 1& 1& 0& 1& 0 \\
 0& 1& 0& 0& 0& 0 \\
 2& 0& -1& -1& 0& -1
\end{array} \right] \cdot P^{-1}&
 \gamma^{\pm}_{10} = \pm P\cdot   \left[\begin{array}{cccccc}
1& 0& -1& 0& 0& 0 \\
 0& 0& 0& 0& 0& 1 \\
 2& -1& -1& 0& -1& 0 \\
 0& 0& -1& 0& 0& 0 \\
 2& 0& -1& -1& 0& -1 \\
 0& 0& 0& 0& 1& 0
 \end{array} \right] \cdot P^{-1}\\[0.50in]
\gamma^{\pm}_{11} = \pm P\cdot   \left[\begin{array}{cccccc}
-1& 0& 1& 0& 1& 0 \\
 2& -1& 0& -1& 0& -1 \\
 0& 0& 1& 0& 0& 0 \\
 -2& 1& 1& 0& 1& 0 \\
 -2& 0& 1& 0& 1& 1 \\
 0& 0& 0& 0& 1& 0
\end{array} \right] \cdot P^{-1}&
 \gamma^{\pm}_{12} = \pm P\cdot   \left[\begin{array}{cccccc}
3& -1& -1& -1& -1& -1 \\
 2& 0& -1& 0& -1& -1 \\
 2& -1& -1& 0& -1& 0 \\
 2& -1& 0& -1& -1& 0 \\
 2& -1& 0& -1& 0& -1 \\
 2& 0& -1& -1& 0& -1
 \end{array} \right] \cdot P^{-1}
\end{array}
\label{12xformmats.eq}
\end{equation}
\end{scriptsize}
Then, the entire 5-crosscap surface or its genus-5 double cover can be
constructed piece by piece starting from a single pentagon $F_{1}$ and
then transforming by $\gamma^{\pm}_{i}$.

\begin {figure}[htb!]
\centering
\mbox{\psfig{width=2.5in, figure=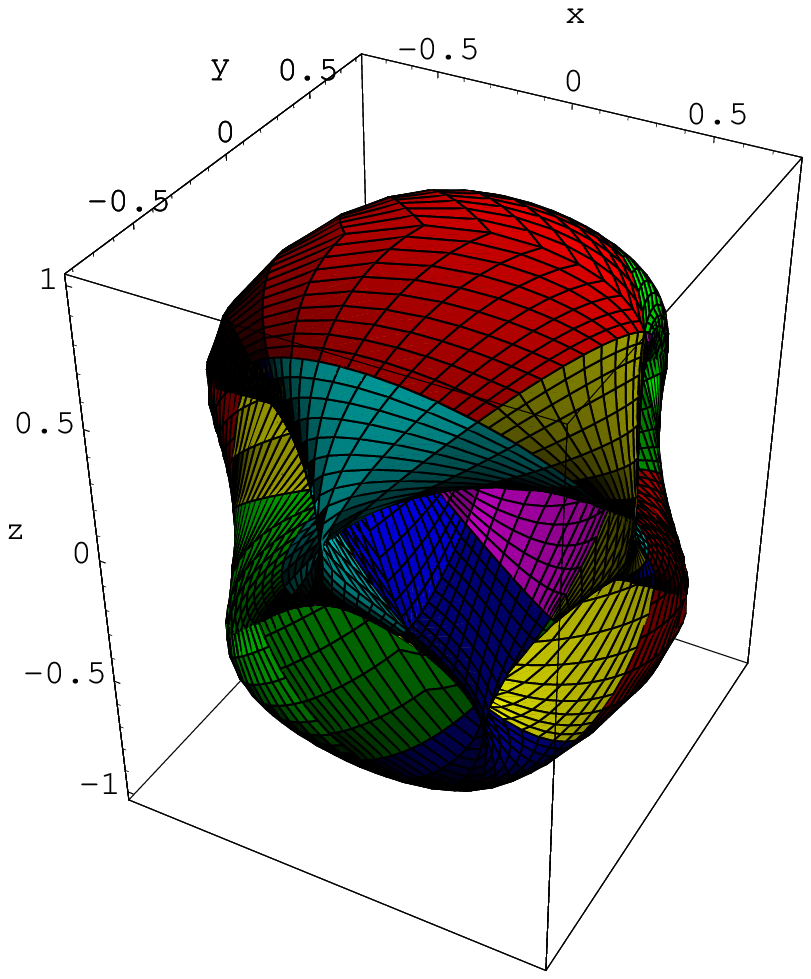}} \hspace{0.1in}
\mbox{\psfig{width=2.5in, figure=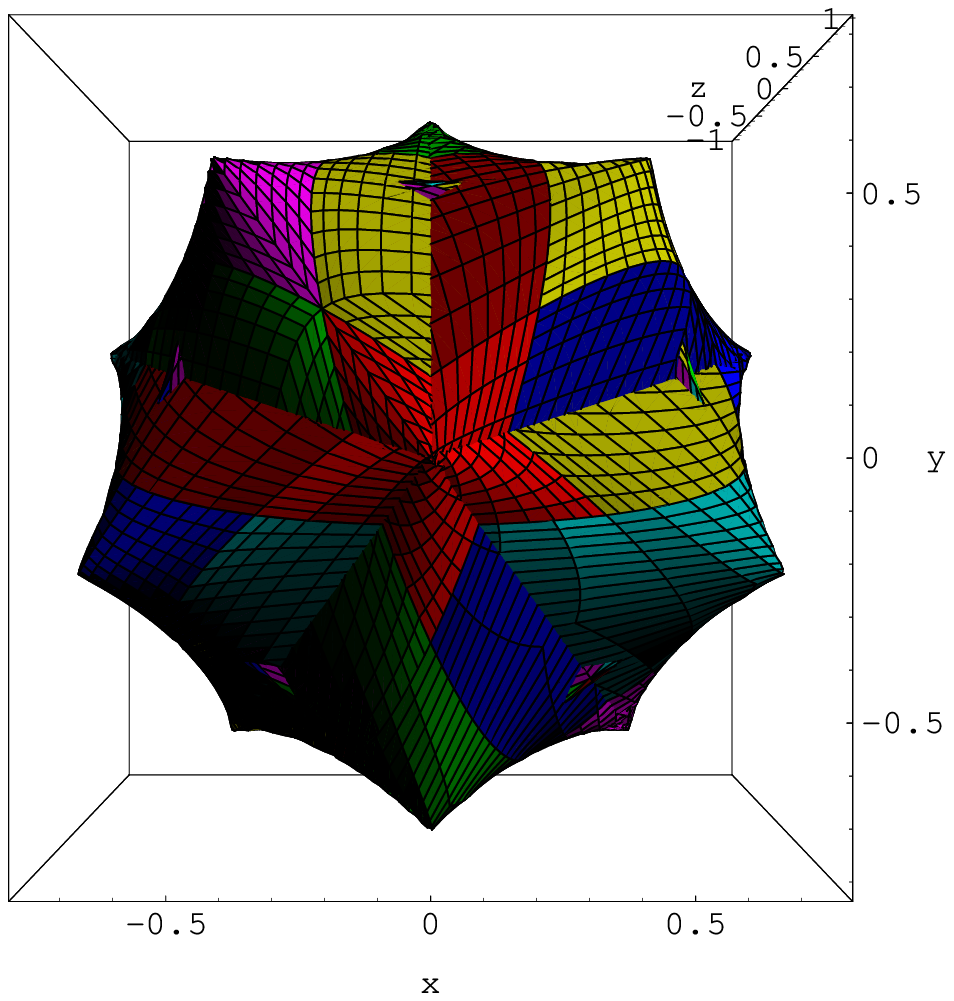}}
\caption [] {Projections from $\RP{5}$ to $\Sphere{5}$ of the double cover
of the 12 blown-up pentagons forming a 5-crosscap dodecahedron. The
first projection emphasizes the smooth nature of the overall surface,
while the second projection emphasizes the pentagonal structure.}
\label {b5rp5flat.fig}
\end {figure}

In Figure \ref{b5rp5flat.fig}, we plot a pair of projections of the 24
surface patches $\gamma^{\pm}_{i} \cdot F_{1}(s,t)$ from $\Sphere{5}$
in $\mathReal^{6}$ to $\mathReal^{3}$. These are global solutions of the
5-point cross-ratio constraints with diametrically opposite copies of
each of the 12 pentagons forming the genus 5 double cover.

\paragraph{$\mathbf B_{5}$ Veronese Map.}
These vertices and the polygonal faces of the dodecahedron inscribed
on the $B_{5}$'s ``real'' integration manifold, a sphere with five
crosscaps, can be compactly embedded for visualization purposes using
a quadratic form that is a  straightforward generalization of
the Veronese surface parameterization.  Given the homogeneous $\RP{5}$
variables $r = (r_0,\,r_1,\,r_2,\,r_3,\,r_4,\,r_5)$
above, we can construct an $\mathReal^{21}$ embedding as
\begin{eqnarray*}
\begin{array}{ccccccccccc}
w_{1} & = & c r_0 r_1 &&
w_{2} & = & c r_0 r_2&&
w_{3} & = & c r_0 r_3 \\
w_{4} & = & c r_0 r_4 &&
w_{5} & = & c r_0 r_5 &&
w_{6} & = & c r_1 r_2 \\
w_{7} & = & c r_1 r_3 &&
w_{8} & = & c r_1 r_4 &&
w_{9} & = & c  r_1 r_5 \\
w_{10} & = & c r_2 r_3 &&
w_{11} & = & c r_2 r_4 &&
w_{12} & = & c r_2 r_5 \\
w_{13} & = & c r_3 r_4 &&
w_{14} & = & c r_3 r_5&&
w_{15} & = & c r_4 r_5 \\
w_{16} & = & (r_{0})^{2} &&
w_{17} & = & (r_{1})^{2} &&
w_{18} & = & (r_{2})^{2} \\
w_{19} & = & (r_{3})^{2} &&
w_{20} & = & (r_{4})^{2} &&
w_{21} & = & (r_{5})^{2}  \ .
\end{array}
\end{eqnarray*}
This map is constructed to lie on the sphere $ \sum_{i} (w_{i})^{2} = 1$
when $c= \sqrt{2}$ and the homogeneous coordinates are normalized
to obey $\sum_{k} (r_{k})^{2}=1$.  Note that the analog of the Steiner
Roman Surface immersion projecting $\RP{2}$  into $\mathReal^{3}$ is achieved
by selecting the variables $(w_{1},\,\ldots,\,w_{15})$ mapping
 $\RP{5}$  into $\mathReal^{15}$.

An alternative, but less symmetric,  $\mathReal^{18}$ embedding is
\begin{eqnarray*}
\begin{array}{ccccccccccc}
w_{1} & = & 2 r_0 r_1 &&
w_{2} & = & \sqrt{2} r_0 r_2&&
w_{3} & = & \sqrt{2} r_0 r_3 \\
w_{4} & = & \sqrt{2} r_0 r_4 &&
w_{5} & = & \sqrt{2} r_0 r_5 &&
w_{6} & = & \sqrt{2} r_1 r_2 \\
w_{7} & = & \sqrt{2} r_1 r_3 &&
w_{8} & = & \sqrt{2} r_1 r_4 &&
w_{9} & = & \sqrt{2}  r_1 r_5 \\
w_{10} & = & 2 r_2 r_3 &&
w_{11} & = & \sqrt{2} r_2 r_4 &&
w_{12} & = & \sqrt{2} r_2 r_5 \\
w_{13} & = & \sqrt{2} r_3 r_4 &&
w_{14} & = & \sqrt{2} r_3 r_5&&
w_{15} & = & 2 r_4 r_5 \\
w_{16} & = & (r_{0})^{2} - (r_{1})^{2} &&
w_{17} & = & (r_{2})^{2} - (r_{3})^{2} &&
w_{18} & = & (r_{4})^{2} - (r_{5})^{2} \ ,
\end{array}
\end{eqnarray*}
which also lies on the sphere $ \sum_{i} (w_{i})^{2} = 1$
when $\sum_{k} (r_{k})^{2}=1$.

\begin {figure}[htbp!]
\centerline{\psfig{width=3.25in, figure=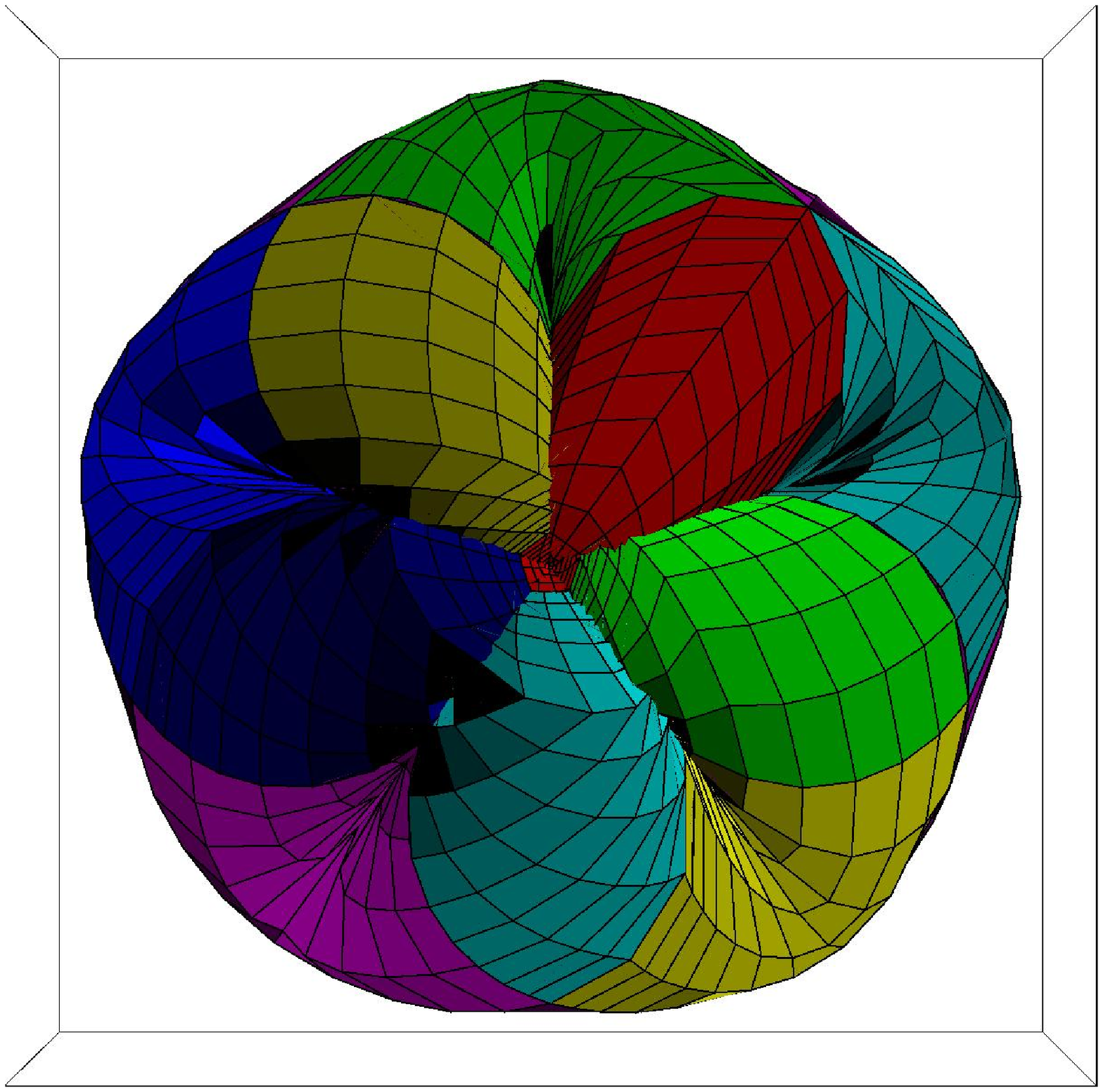}}
\centerline{\psfig{width=3.25in,figure=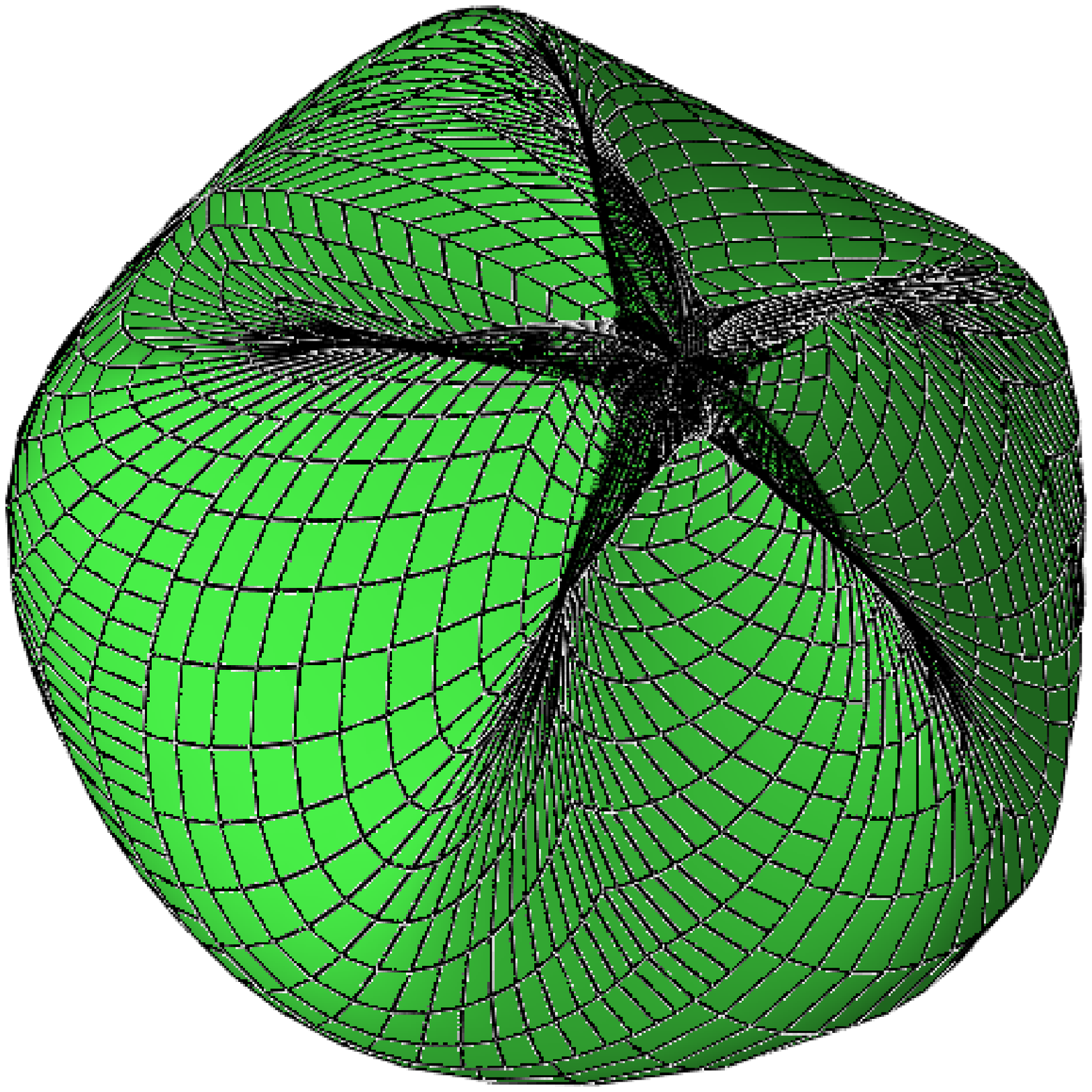}}
\caption [] {Projections of the 5-crosscap surface embedded in
  $\mathReal^{21}$ using  the quadratic map.
Above: the (1,2,5) projection color coded by pentagon.
Below: the (1,2,16) projection with shaded surface and grid.
These are roughly the analogs of the projections of the circle
embedded in the projective plane in Figure \ref{b4-double-rp2.fig}.}
\label {b5vero21.fig}
\end {figure}

Projections of the (non-double-covered) 5-crosscap surface can at last
be drawn using  these quadratic maps, and typical results are
shown in Figure \ref{b5vero21.fig}.


\subsection{The $\mathbf B_5$ Pochhammer Contour}
 \label{sec:VisPoch}

\begin {figure}[htp]
\centering
  \mbox{\psfig{width=8.5cm, figure=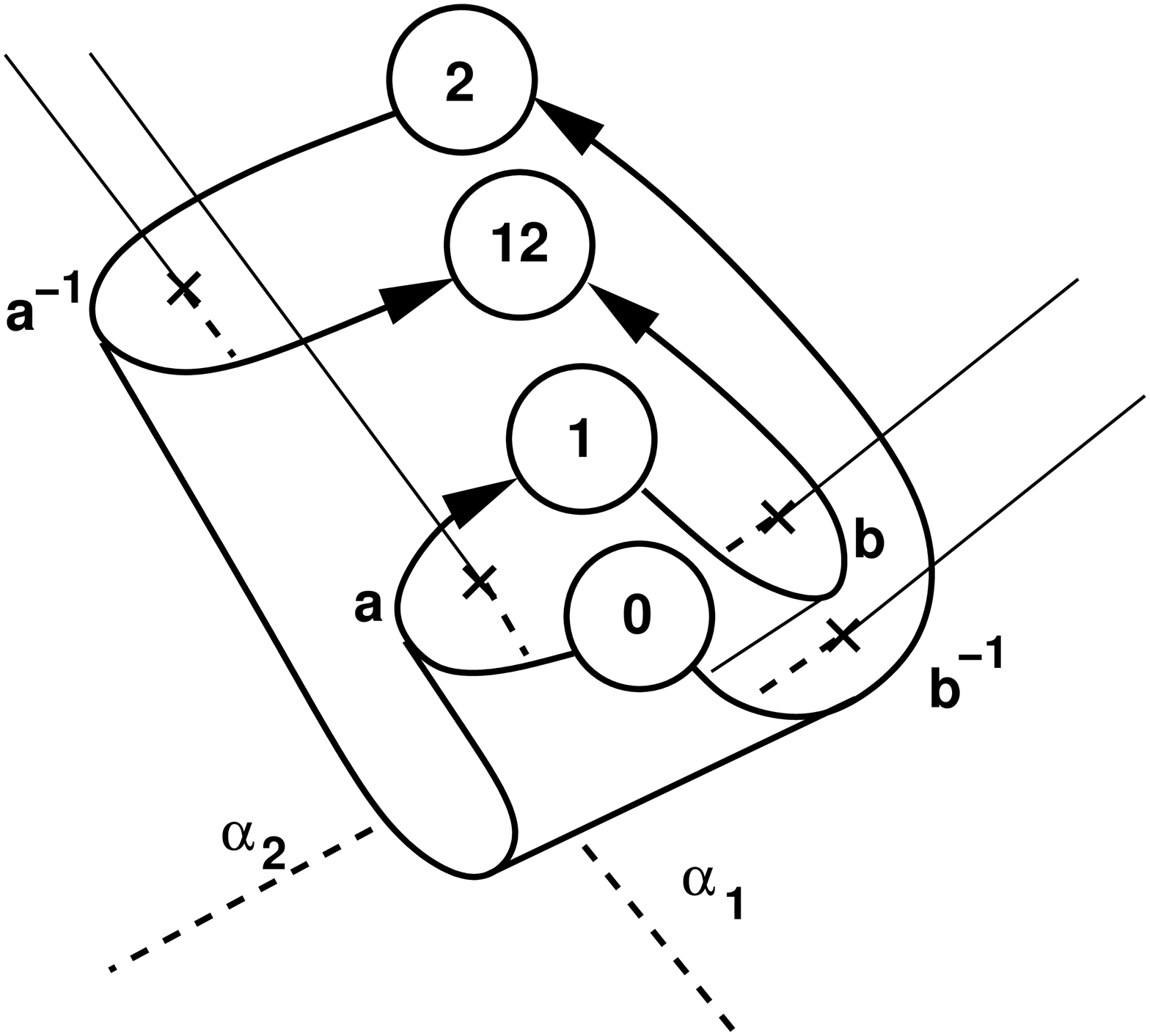}}
\caption [] {The $B_{5}$ ``commutator'' $aba^{-1}b^{-1}$ bounding
a patch that covers one-fifth of each of four different phases,
which for the example branch lines with exponents $\alpha_{1}$ and
 $\alpha_{2}$, are labeled as $\{0,\,1,\,12,\,2\}$.}
\label {B5CommAB.fig}
\end {figure}

\begin {figure}[p]
\centering
  \mbox{\psfig{width=3in, figure=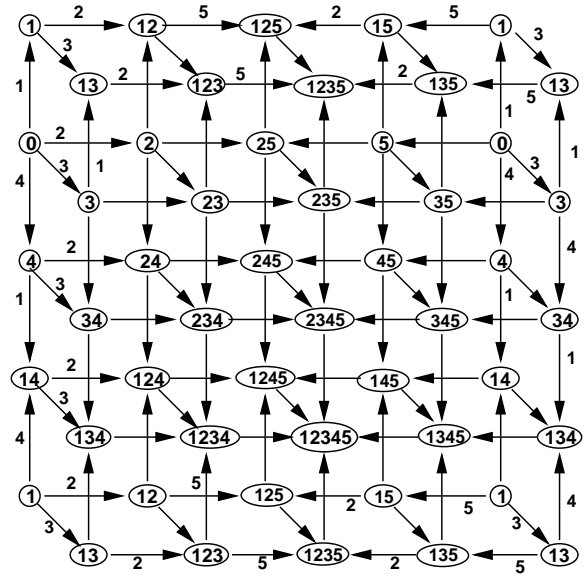}}
\caption [] {The commutator diagram of the $B_5$ Pochhammer surface. }
\label {b5pochform3.fig}
\end {figure}

\begin {figure}[p]
\centering
 \mbox{\psfig{width=3.5in, figure=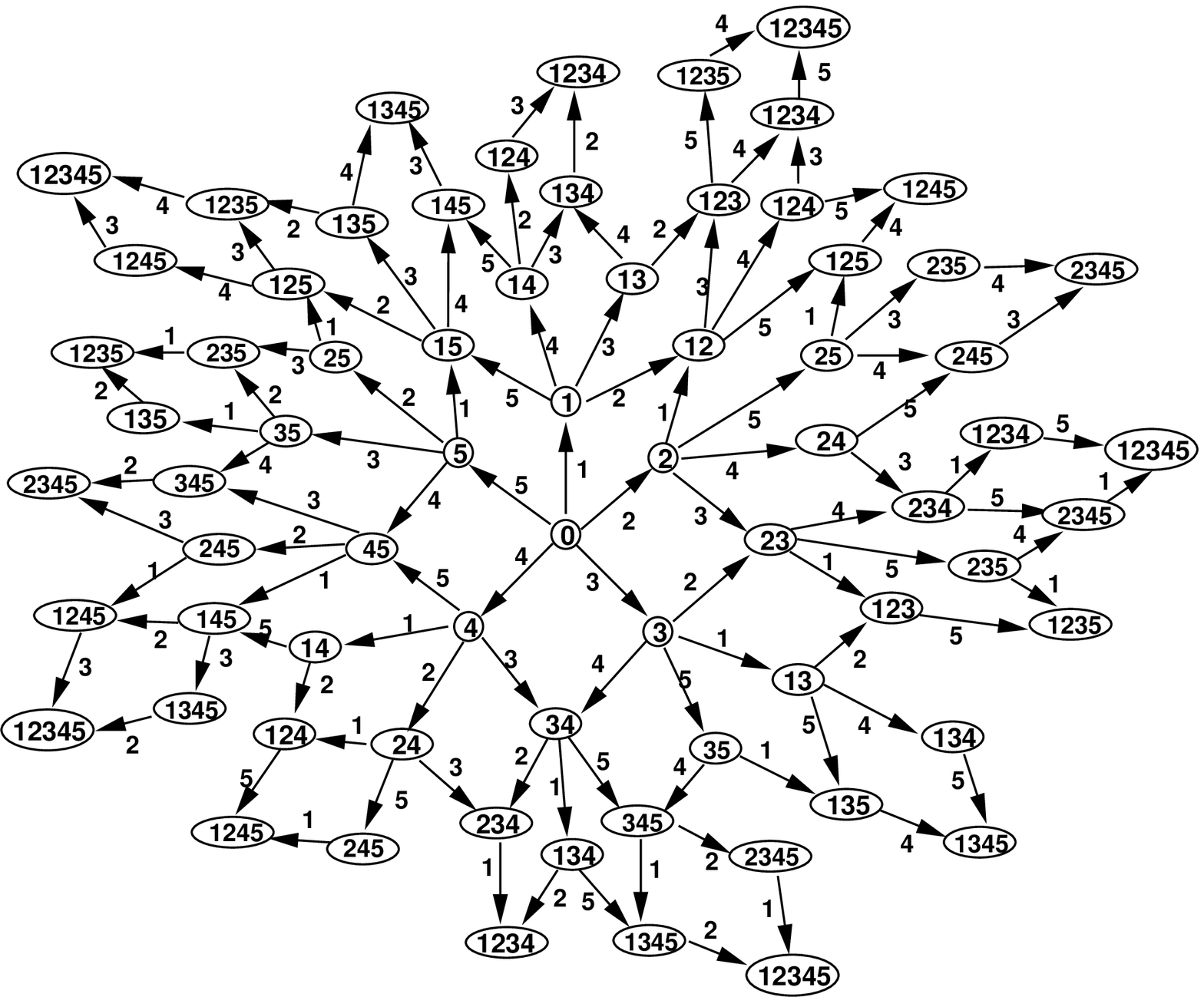}}
\caption [] {The analog of the Pochhammer surface for $B_5$
is a 32-fold cover of a single pentagon, joining each of the
$2^{5}$ possible combinations of the 5 phases
$\{\alpha_{1},\ldots,\alpha_{5}\}$.
This surface can be constructed from (32/4)*5=40 individual
commutator patches.}
\label {b5pochform1.fig}
\end {figure}

\begin {figure}[p]
\centering
 \mbox{\psfig{width=2.5in, figure=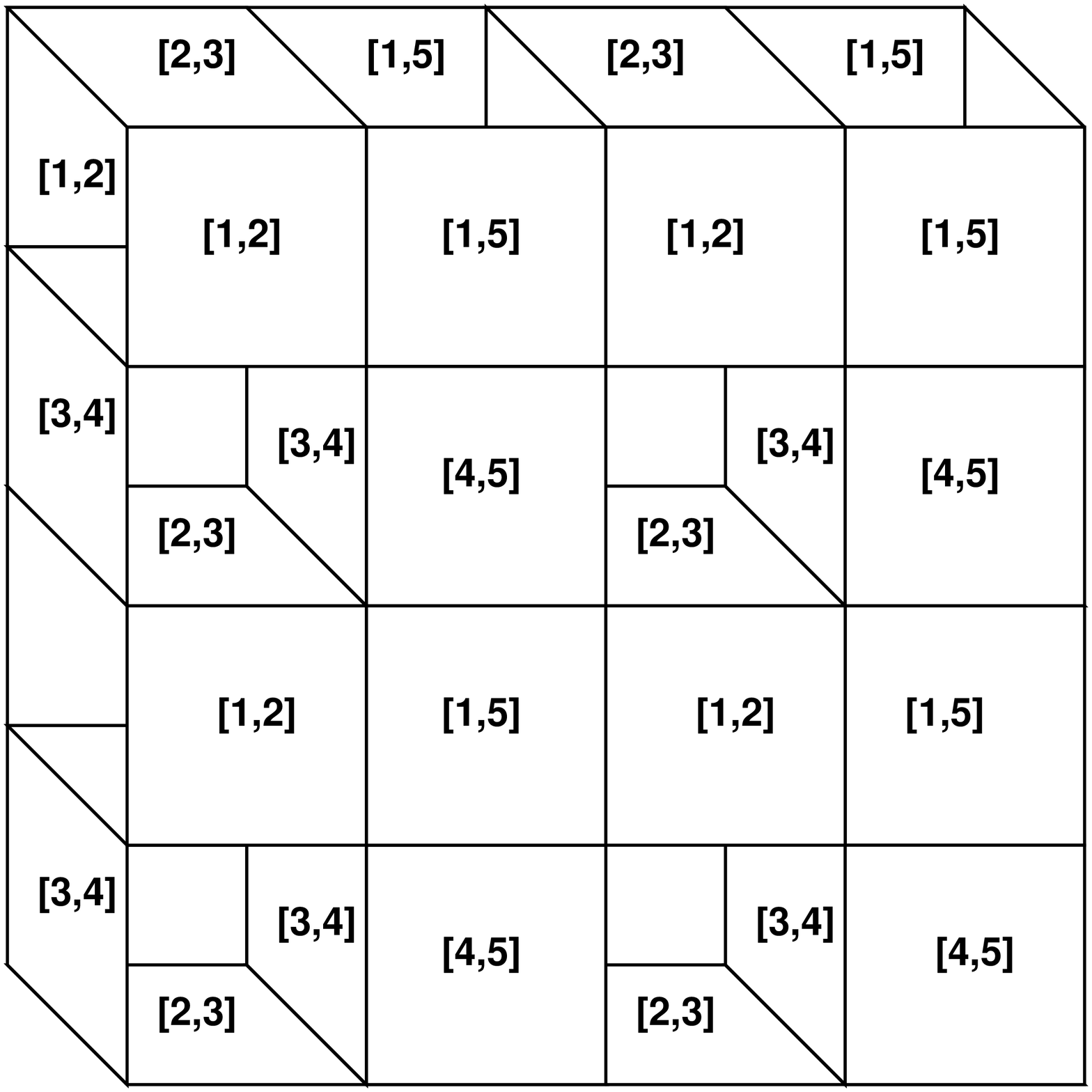}}\\
\caption [] {A global picture of the $B_5$ Pochhammer surface
as an unrolled, thickened torus with four punctures,
showing more clearly the origin of its genus 5 structure.}
\label {b5pochform2.fig}
\end {figure}

Within the domain of a {\it single pentagon\/},
we can now finally begin
to piece together a picture of the global topology of the $B_5$
Pochhammer contour.  This manifold can be drawn explicitly in various
ways by joining together the sets of commutators that eventually
return to the same phase, forming the closed surface; Figure
\ref{B5CommAB.fig} illustrates a single commutator element.    Figure
\ref{b5pochform3.fig} shows the schematic diagram of the full set of
commutators as they return cyclically to the home phase; this diagram
can be unfolded in various ways to show the overall structure, as
illustrated in Figures \ref{b5pochform1.fig} and \ref{b5pochform2.fig}.


\begin {figure}[p]
\centering
\centerline{
  \mbox{\psfig{width=3in,height=4in, figure=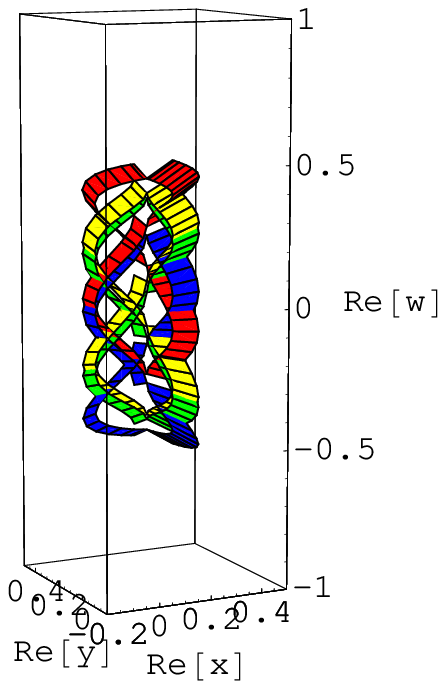}}
  \mbox{\psfig{width=3in,height=4in, figure=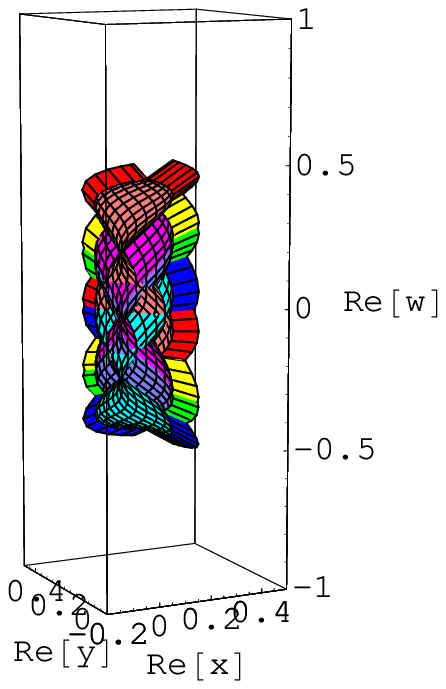}}
}
\centerline{\hspace{0.75in} (a) \hfill (b) \hspace{1.5in}}
\caption [] {(a)
The block of four commutator strips
surrounding one of the five corners of a single $B_5$ Pochhammer
pentagon.
 (b) Filling in the remainder of the surface at one
corner of the pentagon, giving  four ``handkerchief'' surface areas
filling the corresponding branched cover threading its
way around the branch points in the Riemann surface.
}
\label {multikerchief.fig}
\end {figure}

Finally, the explicit algebraic form of the Pochhammer can be
embedded directly in the Riemann manifold of
$\beta_{5}(x,y;\alpha_{1},\ldots,\alpha_{5})$, following the fashion
of Figure  \ref{pochviews.fig} to yield the surfaces shown in
Figure \ref{multikerchief.fig}.  This image shows one-fifth
of the Pochhammer contour covering a set of eight of the  32
total surfaces; sewing together all the corresponding copies
yields the entire surface.

\section{Remarks on the General Case }\label{sec:BNfcn}

\begin {figure}[ht]
\centering
\centerline{
  \mbox{\psfig{width=3in, figure=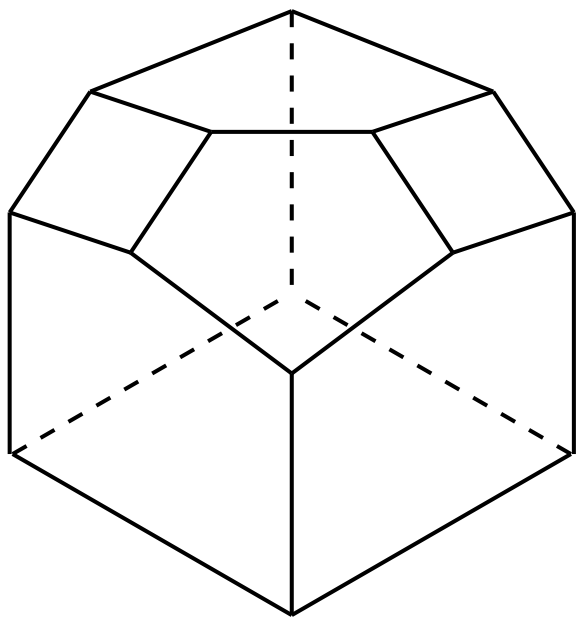}}
}
\caption [] {This nonahedron is the elementary connected component of
the 6-point cross-ratios forming the basis for the analysis of
$B_{6}$; just as 12 pentagons tessellate the 5-crosscap surface,
60 of these nonahedra tessellate the analogous 3-manifold.
}
\label {nonahedron.fig}
\end {figure}

 The affine variety defined by the $N$-point cross-ratio constraints
(\ref{uijconstr.eq}) is of dimension $(N-3)$ and has a natural
decomposition into $(N-1)!/2$ smooth components delineated by the
varieties $u_{ij}=0$.  The $N$-point function $B_{N}$ is initially
defined as an integral of an $(N-3)$-form over a {\it single one\/}
of these components.  Each of the $(N-1)!/2$ components is an {\it
$N(N-3)/2$-polytope\/} --- its $(i,j)$-th face is on the projective
{\it hyperplane\/} given by $u_{ij}=0$; these are not in general
regular polytopes, but reflect the existence of various  poles that
correspond to multiparticle combinations in elementary spinless
string theory.  Table \ref{components.tbl} summarizes for low $N$
the number of cross-ratio variables appearing in the standard
constraints, which is also the number of faces of the polytope
defining a single component, along with the total number of
components.  These polytopes have an exact and previously
unsuspected correspondence with the Stasheff associahedra
\cite{StasheffHSpaces,StasheffTAMS1963,Stasheff2004}, in all dimensions.  Each of the
$B_{6}$ components, for example, is a nonahedron, as pictured in
Figure \ref{nonahedron.fig};  this structure is described in detail
by Devadoss \cite{Devadoss1999}, who also gives, for example, a
tessellation of the moduli space $\overline{\mathcal{M}}^{6}_{0}(\mathReal)$
tiled by 60 nonahedral associahedra.  Our work seems to indicate
that the moduli spaces $\overline{\mathcal{M}}^{N}_{0}(\mathReal)$ studied
by Devadoss can also be viewed as the space of $N$-point
cross-ratios with the tessellations we have described in this paper.

\begin{table}[t!]\centerline{
\begin{tabular}{|r|c|c|} \hline
 \raisebox{-1.5ex}{\rule{0in}{4ex}}$N$ & $\frac{N(N-3)}{2}\, = \,\RP{}$ dim. & %
     $\frac{(N-1)!}{2}$ = no.\  components \\ \hline
 4 &  2 &  3 \\
 5 &  5 &  12\\
 6 &  9 &  60\\
 7 &  14 &  360\\
8 &  20 &  2520\\
 9 &  27 &  20160\\
 10 &  35 &  181440\\
\hline
\end{tabular}}
\caption[]{The first column is the value of $N$ for a given function
  $B_{N}$.  The second column gives the dimension of the projective
  space that implements the blow-up in cross-ratio coordinates; this
  is the same as the number of faces of the polytope defined by the
  natural $u_{ij}=0$ boundaries of the integration region for one
  component  using the cross-ratio coordinates; these are not
  necessarily regular polytopes. The third column is the total number
  of components, i.e., the number of polytopes that fit together to
  give the analogs of the 5-crosscap dodecahedron for $B_{5}$.\\
  \rule{5.375in}{.01in} }
\label{components.tbl}
\end{table}

We conjecture that the real integral form of $B_{N}$ can 
always be expressed alternatively by a Pochhammer-like contour
integral in the corresponding smooth complex algebraic variety $M^{c}$ of
dimension $(N-3)$ in $\CP{N(N-3)/2}$. The contour is a real
$(N-3)$-dimensional submanifold and is obtained by wrapping copies of
the $N(N-3)/2$-polytope integral domain properly around the branch
hyperplanes where its faces are located.  Notice that it is fairly
easy to see, by the description above and (\ref{BNintegral.eq}), that
the branch hyperplane at each face -- say, face $(i,j)$ -- is of
complex codimension 1 in $M^{c}$ and, when folded around it, the {\it
phase\/} of the integrand in the lift to the Riemann covering sheaf
changes by $\pm \alpha_{ij}$.  By this mechanism, (\ref{B5ratio.eq})
should generalize in an obvious way.

\section*{Acknowledgments}
This research was supported in part by NSF grant numbers CCR-0204112
and IIS-0430730.  AJH is grateful to Tullio Regge for his early
encouragement and interest in this problem.   Special thanks are due
to Charles Livingston, Philip Chi-Wing Fu, and Sidharth Thakur for
advice, insights, and assistance with graphics tools.  We also thank
James Stasheff for introducing us to the literature on associahedra.

%
%

\begin{small}
\bibliographystyle{plain}
\bibliography{b5poch}
\end{small}

\end{document}